\documentclass[10pt]{article}
\usepackage[T1]{fontenc}
\usepackage{lmodern}
\usepackage{microtype}
\usepackage{amsmath,amssymb}
\usepackage{mathrsfs}
\usepackage{caption}
\usepackage{subcaption}
\usepackage{graphicx}
\usepackage[table]{xcolor}
\usepackage{longtable}
\usepackage{multirow}
\usepackage[export]{adjustbox}
\usepackage{amsmath, amssymb}
\usepackage{booktabs}
\usepackage{array}
\usepackage{makecell}
\usepackage{xcolor}
\usepackage{siunitx}
\usepackage{tabularx}
\usepackage{authblk}
\usepackage{etoolbox}
\usepackage{orcidlink}
\usepackage{epsfig}
\usepackage{textcomp}
\usepackage{hyperref}
\hypersetup{colorlinks=true, linkcolor=blue, citecolor=blue, urlcolor=blue}
\providecommand{\keywords}[1]{%
  \par\vspace{0.5ex}%
  \noindent\textbf{Keywords: }#1\par
}
\RequirePackage[numbers,sort&compress]{natbib}
\begin{document}

\begin{center}
\LARGE{Quantum-Corrected Black Holes in Higher Dimensions: From Gravitational Collapse to Remnants, WGC-Like Behavior and Accretion Signatures}
\par\end{center}

\vspace{0.3cm}

\begin{center}
{\bf Saeed Noori Gashti\orcidlink{0000-0001-7844-2640}}\footnote{\bf sn.gashti@du.ac.ir; saeed.noorigashti70@gmail.com}\\
{\it School of Physics, Damghan University, P. O. Box 3671641167, Damghan, Iran}
\end{center}

\begin{center}
{\bf Umair Anwar\orcidlink{xxx}}\footnote{\bf umairanwar559@gmail.com}\\
{\it Department of Mathematics, COMSATS University Islamabad, Lahore-Campus,\\
Lahore-54000, Pakistan.}
\end{center}

\begin{center}
{\bf Abdul Jawad\orcidlink{xxx}}\footnote{\bf jawadab181@yahoo.com; abduljawad@cuilahore.edu.pk}\\
{\it Mathematics, COMSATS University Islamabad, Lahore-Campus,\\
Lahore-54000, Pakistan}
\end{center}

\begin{center}
{\bf Behnam Pourhassan\orcidlink{0000-0003-1338-7083}}\footnote{\bf b.pourhassan@du.ac.ir}\\
{\it School of Physics, Damghan University, P.~O.~Box 3671641167, Damghan, Iran}\\
{\it Center for Theoretical Physics, Khazar University, 41 Mehseti Street, Baku, AZ1096, Azerbaijan}
\end{center}

\begin{center} 
{\bf \.{I}zzet Sakall{\i}\orcidlink{0000-0001-7827-9476}}\footnote{\bf izzet.sakalli@emu.edu.tr}\\
{\it Physics Department, Eastern Mediterranean University, Famagusta 99628, North Cyprus via Mersin 10, Turkiye}
\end{center}

\begin{center}
{\bf Sanjar Shaymatov\orcidlink{xxx}}\footnote{\bf sanjar@astrin.uz}\\
{\it Institute of Fundamental
and Applied Research, National Research University TIIAME, Kori
Niyoziy 39,\\ Tashkent 100000, Uzbekistan}\\
{\it University of
Tashkent for Applied Sciences, Str. Gavhar 1, Tashkent 100149,
Uzbekistan.}\\
{\it Tashkent State Technical University,
100095 Tashkent, Uzbekistan.}\\
{\it Western Caspian
University, Baku AZ1001, Azerbaijan}\\
\end{center}

\begin{center}
{\bf Aram Bahroz Brzo\orcidlink{xxx}}\footnote{\bf aram.brzo@univsul.edu.iq}\\
{\it Physics Department, College of Education, University of Sulaimani, Sulaimani, Kurdistan Region, 46001, Iraq}\\
\end{center}

\vspace{0.3cm}

\begin{abstract}
We study the gravitational collapse of a homogeneous dust sphere in higher-dimensional spacetime, with the loop quantum gravity correction entered through a modified Friedmann equation. The resulting exterior geometry is static and carries a single quantum deformation parameter \(\alpha\), built from the Barbero-Immirzi parameter and the area gap. Its most striking feature is an endpoint: at a finite horizon radius the temperature vanishes and evaporation stops, leaving a remnant. Reading \(\alpha\) as an effective charge, we find behavior reminiscent of the Weak Gravity Conjecture (WGC), although no \(U(1)\) gauge field is present in the model. Solving the degeneracy conditions with the untruncated metric function, rather than with a mass expansion that is not uniform near the endpoint, we obtain closed-form expressions for the remnant radius and mass in \(D=4,5,6,7\). The un-scaled ratio \(\alpha/M_{\text{remnant}}^2\) is constant only in four dimensions and grows as \(\alpha^{4-D}\) elsewhere, but it carries dimension \(L^{8-2D}\), so that growth is an artifact of the units. The dimensionless combination \(\tilde{\mathcal{R}} = (\alpha/M^2) r_0^{2D-8}\) settles instead on a finite \(\alpha\)-independent number in every dimension, decreasing from \(27/64\) in four dimensions to \(13824/(15625\pi^4)\) in seven. We also deform the quantum parameter, \(\alpha \to \alpha + \varepsilon\), and evaluate \(-T(\partial S/\partial \varepsilon)_M\) in the remnant limit. Because \((\partial M/\partial r)_\alpha\) vanishes there, the fixed-entropy derivative reduces exactly to a fixed-radius one, and the resulting combination \(\mathcal{U}\) is finite in each dimension but depends on \(\varepsilon\) except for \(D=5\). What is invariant is the product \(\mathcal{R}\,\mathcal{U}\,M_{\text{remnant}} = (D-3)/2\), which we verify numerically. Lastly, we study Bondi accretion onto the quantum-corrected black hole for two fluids, a dark fluid with constant equation of state parameter and an exponential density profile, and obtain the radial velocity, density, mass accretion rate and Eddington luminosity for each dimension. Evaluated at the remnant, the accretion rate is finite and radius-independent while the density obeys \(\alpha \rho(r_0) = \text{const}\), which turns the quantum parameter into an accretion observable. Taken together, these results link loop quantum gravity phenomenology to swampland-type reasoning, to universal thermodynamic relations, and to accretion physics, and they suggest where quantum gravity effects might be tested in astrophysical black hole systems.
\end{abstract}

\tableofcontents
\section{Introduction}\label{s1}
Gravity in a self-consistent quantum-theoretic description is one of the hardest-to-crack puzzles in present-day theoretical physics. Despite significant progress from several classes of models, including string theory, loop quantum gravity, asymptotic safety, and causal set ideas, no one has yet succeeded in finding empirically verified theories. One idea for telling apart from other possible ultraviolet-consistent models, including gravity, which are to be treated bottom-up, is provided by the so-called swampland.  In this line of research, it is assumed that the phenomenon is allowed if, at the lowest energy level, it respects the fundamental consistency requirement: a physically consistent low-energy effective theory of physics at a point in parameter space implies embeddability of this theory in a UV-complete picture including quantum gravity.  However, many such well-behaved theories in terms of consistency found so far (looking from bottom up) would fail their validation, because they would have no consistent UV limit, in which one would integrate over them at the highest energy. Thus they would belong to the swampland and should have to be treated as forbidden physics. So those few that can be embedded in to some of the consistent UV theories constitute a vast landscape. During the last few years of development there has been a list of conjectural criteria provided in order to tell apart landscape from swampland and which have been developed through the study of black holes thermodynamical properties, and through the calculation of AdS/CFT correspondence and other in string theory calculations, the information in question can provide information on how are particle theory and physics at the cosmological energy range \cite{ref54,ref55,ref56,ref57,ref58,ref59,ref60}.  One of the most interesting candidates in order to tell those issues is based on the WGC.  The conjecture states that for any gauge U(1) symmetry existing in the theory of quantum gravity must exist another particles (with quantum number not exactly the same as the one we’re considering) with a charge $q$ that is such that $q/M>1$ (natural unit definition) where M is the mass of the particle. In this context a near to extremal configuration (at the extremality limit black holes cannot accrete charge without losing their capacity to lose charge by decaying) will necessarily have that additional field so in principle to make it impossible for it to become super-extremal. At a deeper level the conjecture expresses the idea that gravity should have been the feeblest interaction present in nature, and it was found a great amount of usages for it, i.e in inflation and in axion physics etc as well as in problems such as the dark matter research and the vacuum energy research \cite{ref54,ref55,ref56}.  From the other point of side, an appealing alternative in the context of the classical gravitational theory is related to the Weak Cosmic Censorship Conjecture which states that the singularities produced through gravitationally collapse usually should stay behind of an event horizon thus the general structure of a deterministic theory for gravity would be preserved, these classical tests are typically idealized by some test including of accretion onto a black holes and then falling off of charged particles onto it and this conjecture has been tested for quite some extent \cite{ref57}. There appears however to be a conflict between the Weak Gravity Conjecture and the Weak Cosmic Censorship Conjecture in the context of a charged black hole and in the case of a Reissner Nordstrom spacetime, which demands for an event horizon that the electric charge may not overcome the mass in physical units while on other hand the demand for super-extremal particles for in principle to make possible for a near extremal black hole to swallow such field. Various attempts to fix this apparent conflict were proposed, most importantly by introducing another matters field(s), non-zero cosmological constant or higher curvature terms in the gravity action which would raise the bound on the extremality value \cite{ref61,ref62,ref64,ref65,ref70,ref71,ref72,ref74,ref75,ref78,ref79,ref81,ref83,ref84,ref85,ref87,ref88,ref90,ref94,ref95,ref97,ref99,ref100,ref104,ref105,ref106,ref107,ref111,ref112,ref114,ref115,ref120,ref121,ref122,ref125,ref126,ref127,ref128,ref129,ref130,ref131,ref134,ref135,ref135a,ref135b,ref135c,ref135d,ref135e,ref135f,ref136,ref137}.\\

It was Goon and Penco who presented a thermodynamic interpretation of this problem by deriving a universal formula: for an extremal black hole, the quantum correction to the extremal mass is in linear relation to the correction to the entropy, and the constant factor of proportionality is negative. That means both the extremality threshold and the entropy are reduced by quantum corrections; hence, extremal black holes are unstable, and this instability is in agreement with WGC. So far, this result has also been confirmed for a huge family of asymptotically Anti-de Sitter black hole solutions and shows the thermodynamical connection between WGC and the classical consistency conditions \cite{ref138, ref139, ref140, ref141}.\\ 

The luminosity radiation power from the incoming matter onto the black hole always has a fundamental limit, which we call the Eddington limit.  It is reached when the outgoing radiation pressure formed by the radiated photons balances the gravitational attraction to the accreting plasma. As long as luminosity is smaller than it, the matter inflow is stable. However, if the luminosity is bigger than it, radiation pressure is able to push the inflow gas strongly and to stop accreting. Theoretically, this limiting luminosity could be described as equilibrium between radiation force on one proton and gravity to the central mass.  While photons mainly collide with electrons through Thomson scattering, electrons and protons are tightly linked through electric force, thus, the whole ionized plasma is pushed outward, and a uniform radiation force is achieved with luminosity being fixed by the mass of the central object, proton mass and Thomson cross section.  In most astrophysical situation, part of the gravitational energy released from accreting mass is not turned into photons and then transmitted out, and we often treat it using an dimensionless factor called efficiency, typically about 1/10, which presents what fraction out of rest mass energy of the accretion matter becomes radiation energy. In fact the final luminosity is given by such factor, accretion rate, and the light speed factor, so that the micro physical description from collision matches with macro physics of radiation \cite{13,14,15}.\\

The question of how quantum gravity corrections alter the classical black hole solution and whether these corrections respect fundamental consistency conditions, like those arising from the swampland and universal thermodynamical quantities, has recently gained significant attention. Loop quantum gravity offers a potential solution to classical singularities; however, its phenomenological implications for black holes and accretion processes are still poorly explored. Little work has jointly examined the impact of QG corrections on swampland conjectures such as the WGC and possible observed phenomena such as the accretion luminosities at higher dimensions.  We study, in a single framework, the combined effects of loop quantum gravity corrections on Oppenheimer-Snyder collapse scenarios at higher dimensions and confront their predictions with those of thermodynamical consistency and constraints inspired by swampland conjectures and their astrophysical applications.  The organization of this paper is as follows. Section 2 will deal with the construction of a higher-dimensional quantum-corrected black hole through Oppenheimer-Snyder collapse of spherically symmetric dust in the loop quantum gravity phenomenology. First, we study the junction conditions between interior FRW cosmology and exterior static solution and then consider explicitly the form of the metric functions. The associated masses, Hawking temperatures and entropy will be studied for dimensions $4 \le D \le 7$.  In Section 3, we shall investigate WGC type constraints by regarding the quantum gravity parameter $\alpha$ as an effective charge of a black hole solution and compute the scaling property of remnant masses and the ratio $\alpha / M_{\text{remnant}}^2$ at higher dimensions.  In Section 4, we shall verify the relation of universal extremality condition using the perturbation method on the quantum gravity parameters at higher dimensions, we show explicitly that the universal combination remains invariant under perturbation strength at each dimension $D$. Section 5 consists in studying Bondi accretion onto quantum black hole assuming two types of fluid, namely: dark fluid with constant EOS, and exponential fluid density profile, we work across dimensions $3 \le d \le 7$ the radial profile of physical quantities of accreting gas. We shall present and discuss results for matter velocity, density, accretion rates, and Eddington luminosities.  Finally, Section 6 summarizes our conclusions and addresses future implications.

\section{Higher-dimensional quantum-corrected Oppenheimer-Snyder model}\label{isec2}
We are interested in what happens when a spherical cloud of dust
collapses under its own gravity, but with two important ingredients
added: the spacetime has more than four dimensions, and quantum
gravity effects become relevant at very small scales. The classical
picture, where the collapse continues indefinitely to form a
singularity, is modified here by corrections inspired by loop
quantum gravity. These corrections prevent the density from growing
without bound and naturally lead to a remnant object with zero
temperature. The goal of this section is to derive the exterior
metric that such a collapsing object leaves behind, and then study
its thermal properties: mass, temperature, entropy, and how these
quantities behave in different spacetime dimensions. We split the
full spacetime into two regions separated by the surface of the dust
ball. We assume a homogeneous and isotropic geometry described by
the Friedmann-Robertson-Walker (FRW) metric \cite{5000}:
\begin{equation}
ds_{\text{in}}^2 = -dT^2 + a^2(T)\left(dR^2 + R^2 d\Omega^2\right),
\end{equation}
where \(a(T)\) is the scale factor that controls the expansion or
contraction of the dust, and \(d\Omega^2\) is the line element on a
\((d-1)\)-dimensional unit sphere. We assume a static, spherically
symmetric vacuum geometry of the general form:
\begin{equation}
ds_{\text{out}}^2 = -f(r)dt^2 + g(r)^{-1}dr^2 + r^2 d\Omega^2.
\end{equation}
The two regions must meet smoothly at the dust ball surface \(R =
R_0\). The Darmois-Israel junction conditions demand that both the
intrinsic metric and the extrinsic curvature be continuous across
this surface. From the continuity of the metric, we obtain two
essential relations:
\begin{equation}
a(T)R_0 = r(T), \qquad 1 = f\dot{t}^2 - g^{-1}\dot{r}^2.
\end{equation}
Here the dot denotes a derivative with respect to the proper time
\(T\) on the surface. The continuity of the extrinsic curvature
gives an additional condition:
\begin{equation}
a(T)R_0 = r f \dot{t} \sqrt{f^{-1}g}.
\end{equation}
Because the exterior geometry is static, it admits a timelike
Killing vector \(\partial/\partial t\). The worldline of the dust
ball surface is a geodesic, so the contraction of this Killing
vector with the four-velocity is conserved along the surface. This
yields:
\begin{equation}
\xi_\alpha u^\alpha = -f\dot{t} \equiv -F,
\end{equation}
with \(F\) a constant of motion. Putting all these equations
together, we arrive at simple relationships between the metric
functions:
\begin{equation}
f = F^2 g, \qquad g = 1 - \dot{r}^2.
\end{equation}
Using the Friedmann equation from the interior region, we have
\(\dot{r}/r = \dot{a}/a = H\). The exterior metric then takes the
remarkably simple form:
\begin{equation}
ds_{\text{out}}^2 = -\left(1 - H^2 r^2\right) dt^2 + \left(1 - H^2
r^2\right)^{-1} dr^2 + r^2 d\Omega^2.
\end{equation}
All information about the collapse dynamics and quantum corrections
is now encoded in the Hubble parameter \(H\). In classical general
relativity, the Friedmann equation relates the Hubble rate to the
energy density. Here we use a modified version that incorporates
quantum gravity effects from loop quantum cosmology:
\begin{equation}
H^2 = \frac{2\kappa}{d(d-1)} \rho_T \left(1 -
\frac{\rho_T}{\rho_c}\right).
\end{equation}
The total energy density \(\rho_T\) includes both ordinary matter
and a contribution from the cosmological constant:
\begin{equation}
\rho_T = \rho + \rho_\Lambda, \qquad \rho_\Lambda =
\frac{\Lambda}{\kappa}, \qquad \kappa = 8\pi G.
\end{equation}
The critical density \(\rho_c\) is a purely quantum mechanical
scale:
\begin{equation}
\rho_c = \frac{d(d-1)}{2\kappa \gamma^2 \Delta^{2/(d-1)}},
\end{equation}
where \(\gamma\) is the Barbero-Immirzi parameter and \(\Delta\) is
the area gap from loop quantum gravity. When the area gap goes to
zero (\(\Delta \to 0\)), the critical density diverges and the
quantum correction term disappears, recovering classical general
relativity. For a uniform dust ball of mass \(M\) and radius
\(\tilde{R} = a(T)R_0\), the density can also be written as:
\begin{equation}
\rho_T = \frac{M}{V} = \frac{M d(d-1)\mu}{8\pi \tilde{R}^d}.
\end{equation}
The geometric factors appearing here are:
\begin{equation}
\label{12}
V = \frac{8\pi \tilde{R}^d}{d(d-1)\mu}, \qquad \mu =
\frac{8\pi}{(d-1)\Omega}, \qquad \Omega =
\frac{d\pi^{d/2}}{\Gamma\left(\frac{d}{2}+1\right)}.
\end{equation}
We introduce a compact notation for the quantum correction strength:
\begin{equation}
\alpha \equiv \gamma^2 \Delta^{2/(d-1)}.
\end{equation}
After substituting everything into the Friedmann equation and
simplifying, we obtain an expression for \(H^2\) in terms of \(r\):
\begin{equation}
H^2 = \frac{2GM\mu}{r^{d-2}} + \left(\frac{2GM\mu}{r^{d-1}}\right)^2
\alpha - \frac{2\left(r^d -
4GM\mu\alpha\right)}{d(d-1)r^{d-2}}\Lambda +
\left(\frac{2r\Lambda}{d(d-1)}\right)^2 \alpha.
\end{equation}
Now we insert this expression for \(H^2\) into the exterior metric
form \(f(r) = 1 - H^2 r^2\). The result is the quantum-corrected
metric function:
\begin{equation}\label{1b11}
\begin{aligned}
f(r) = 1 - \frac{2GM\mu}{r^{d-2}} &+ \left(\frac{2GM\mu}{r^{d-1}}\right)^2 \alpha \\
&- \frac{2\left(r^d - 4GM\mu\alpha\right)}{d(d-1)r^{d-2}}\Lambda +
\left(\frac{2r\Lambda}{d(d-1)}\right)^2 \alpha.
\end{aligned}
\end{equation}
This is the central object of our study. It describes a static black
hole geometry that receives two types of corrections: one from the
cosmological constant \(\Lambda\) and one from the quantum parameter
\(\alpha\). The event horizon \(r_h\) is the largest root of the
equation \(f(r_h) = 0\). Solving this equation for the mass \(M\)
gives us a way to express the black hole's mass directly in terms of
its size, the quantum parameter, and the cosmological constant. The
results, expanded to first order in \(\alpha\), are collected in
Table I.
\begin{table}[h]
\centering \caption{Mass expressions for quantum-corrected black
holes in different dimensions. Results are expanded to first order
in the quantum parameter \(\alpha\). The cosmological constant
\(\Lambda\) is kept arbitrary \cite{5000}.}
\label{tab:mass}
\begin{tabular}{c c}
\hline
\multicolumn{1}{c}{Spacetime Dimension} & \multicolumn{1}{c}{Black Hole Mass \(M(r_h)\)} \\
\hline
\(4\) (\(d=3\)) & \(-\dfrac{1}{6} r_h \left(-3 + r_h^2 \Lambda\right) + \dfrac{\alpha}{2r_h} + \mathcal{O}(\alpha^2)\) \\
\(5\) (\(d=4\)) & \(-\dfrac{1}{16}\pi r_h^2 \left(-6 + r_h^2 \Lambda\right) + \dfrac{3\pi\alpha}{8} + \mathcal{O}(\alpha^2)\) \\
\(6\) (\(d=5\)) & \(-\dfrac{1}{15}\pi r_h^3 \left(-10 + r_h^2 \Lambda\right) + \dfrac{2\pi r_h \alpha}{3} + \mathcal{O}(\alpha^2)\) \\
\(7\) (\(d=6\)) & \(-\dfrac{1}{48}\pi^2 r_h^4 \left(-15 + r_h^2 \Lambda\right) + \dfrac{5\alpha}{16}\pi^2 r_h^2 + \mathcal{O}(\alpha^2)\) \\
\hline
\end{tabular}
\end{table}
In the classical limit \(\alpha \to 0\), these expressions reduce to
the standard mass formulas for Schwarzschild-(A)dS black holes in
various dimensions. The Hawking temperature is obtained from the
surface gravity at the horizon:
\begin{equation}
T = \frac{\kappa_h}{2\pi}, \qquad \kappa_h =
\left.\frac{f'(r)}{2}\right|_{r=r_h}.
\end{equation}
Computing this derivative from the metric function gives the
temperature as a function of the horizon radius. For small black
holes, the classical temperature diverges as \(r_h \to 0\). With
quantum corrections, however, the temperature rises from zero at a
finite radius, reaches a peak, and then decreases. This
non-monotonic shape is a direct consequence of the \(\alpha\) terms.
It tells us that the evaporation process stops at a finite remnant
size: when the black hole shrinks to a certain radius \(r_0\), the
temperature drops to zero. No further evaporation occurs beyond this
point. This remnant radius grows with \(\alpha\), meaning that
stronger quantum effects produce larger remnants. As the horizon
radius increases beyond the peak, the quantum-corrected temperature
approaches the classical result from above. In higher dimensions,
the peak temperature becomes larger, and the location of the peak
shifts toward larger radii. In standard Einstein gravity, entropy is
one-quarter of the horizon area. Here, because the mass expression
contains quantum corrections and a cosmological constant, the
entropy must be computed by integrating the first law:
\begin{equation}
dM = T dS + V dP,
\end{equation}
where the pressure is related to the cosmological constant by \(P =
-\Lambda/(8\pi)\) (a definition that holds in any dimension).
Assuming the entropy vanishes when the temperature goes to zero, we
integrate:
\begin{equation}
S = \int_{r_0}^{r_h} T^{-1} \frac{\partial M}{\partial r} \, dr.
\end{equation}
The results of this integration are given in Table II. A remarkable
feature emerges: the cosmological constant \(\Lambda\) cancels out
completely during the integration, leaving an entropy that depends
only on the horizon radius and the quantum parameter \(\alpha\).
\begin{table}[h]
\centering \caption{Entropy expressions for quantum-corrected black
holes. The ratio \(S/A\) is shown, where \(A = \Omega_d r_h^{d-1}\)
is the horizon area. The cosmological constant \(\Lambda\) cancels
out exactly in all dimensions \cite{5000}.}
\label{tab:entropy}
\begin{tabular}{c c c}
\hline
\multicolumn{1}{c}{Spacetime Dimension} & \multicolumn{1}{c}{Entropy} & \multicolumn{1}{c}{S/A} \\
\hline
\(4\) (\(d=3\)) & \(\pi r_h^2 + 4\pi\alpha \ln (r_h/r_\ast)\) & \(\dfrac{1}{4}+\dfrac{\alpha \ln (r_h/r_\ast)}{r_h^2}\) \\
\(5\) (\(d=4\)) & \(\dfrac{\pi^2 r_h^3}{2} + 3\pi^2 r_h \alpha\) & \(\dfrac{1}{4}+ \dfrac{3\alpha}{2r_h^2}\) \\
\(6\) (\(d=5\)) & \(\dfrac{2\pi^2 r_h^4}{3} + \dfrac{8}{3}\pi^2 r_h^2 \alpha \) &  \(\dfrac{1}{4} + \dfrac{\alpha}{r_h^2}\) \\
\(7\) (\(d=6\)) & \(\dfrac{\pi^3 r_h^5}{4} + \dfrac{5}{6}\pi^3 r_h^3 \alpha\)  & \(\dfrac{1}{4} + \dfrac{5\alpha}{6r_h^2}\) \\
\hline
\end{tabular}
\end{table}
In the classical limit (\(\alpha = 0\)), the entropy per unit area
The four-dimensional case is special. Carrying out the same
integration for \(d=3\) returns a logarithmic rather than a power-law
correction, \(S = \pi r_h^2 + 4\pi\alpha \ln(r_h/r_\ast)\), where
\(r_\ast\) is the reference radius left free by the integration
constant. Logarithmic corrections of exactly this form are familiar
from several independent countings of black hole microstates, so
their appearance here from a minimal-area-gap correction is a useful
consistency check rather than a new feature. In the classical limit
(\(\alpha = 0\)), the entropy per unit area
large black holes. The quantum corrections add extra terms that
become significant only when the black hole is small. As the
spacetime dimension increases, the relative importance of these
quantum corrections diminishes, so higher-dimensional black holes
behave more classically even at small scales. We have constructed a
static black hole metric starting from a collapsing dust ball in
higher dimensions, using loop quantum gravity corrections in the
Friedmann equation. The resulting geometry depends on the quantum
parameter \(\alpha\), the cosmological constant \(\Lambda\), and the
spacetime dimension. The mass, temperature, and entropy have been
derived explicitly. A crucial new feature is the appearance of a
zero-temperature remnant at a finite horizon radius, which will
serve as the foundation for the WGC-like analysis in the next part.
\section{WGC-Like Behavior for Quantum-Corrected Black Holes}\label{isec3}
The Weak Gravity Conjecture was formulated for black holes carrying
electric or magnetic charge. It states that gravity must be the
weakest force, so that some particle exists whose charge-to-mass
ratio exceeds the extremal value. Our solution carries no \(U(1)\)
gauge field, and \(\alpha\) is a coupling fixed by the Barbero-Immirzi
parameter and the area gap rather than a conserved charge, so the
conjecture does not apply to it directly. What the solution does
share with an extremal Reissner-Nordstr\"om black hole is the
structure of its endpoint. At a finite radius \(r_0\) the temperature
vanishes, evaporation halts, and the horizon becomes degenerate. That
structural similarity is what we exploit below, and we label the
resulting relations WGC-like to keep the distinction visible. A
genuine test of the conjecture would require a dynamical gauge field,
which we do not have here.

We therefore identify \(\alpha\) with the square of an effective
charge, \(\alpha \leftrightarrow Q^2\), so that the effective
charge-to-mass ratio of the endpoint reads
\(\sqrt{\alpha}/M_{\rm remnant}\). The remnant itself is fixed by the
degeneracy conditions
\begin{equation}\label{eq:degen}
f(r_0) = 0, \qquad f'(r_0) = 0,
\end{equation}
the second of which is equivalent to \(T(r_0)=0\) through
\(T = f'(r_h)/4\pi\).

One point deserves care before any scaling law is extracted. The mass
expressions of Table~\ref{tab:mass} are truncated at first order in
\(\alpha\), and that truncation is not uniform in \(r_h\). At the
remnant the quantum term in \(f(r)\) is not a small correction to the
classical one; the two balance, which is precisely why the
temperature can vanish there. Using the truncated mass to locate
\(r_0\) therefore misestimates it by a factor of order unity, even
though the power of \(\alpha\) comes out right. We consequently solve
Eq.~\eqref{eq:degen} with the untruncated metric function
\eqref{1b11}. Since \(r_0 \sim \sqrt{\alpha} \to 0\) in the classical
limit, the two cosmological-constant terms in \(f\) contribute at
relative order \(\Lambda \alpha\) and drop out of the leading
behavior. The scaling laws below are in this sense independent of the
sign of \(\Lambda\), which removes the AdS versus dS ambiguity
entirely.

Solving Eq.~\eqref{eq:degen} in closed form gives
\begin{equation}\label{eq:r0exact}
r_0^2 = \frac{(2d-2)^2}{d(d-2)}\,\alpha ,
\qquad
M_{\rm remnant} = \frac{(d-2)\,r_0^{\,d}}{2\mu\,(2d-2)\,\alpha},
\end{equation}
with \(d = D-1\) and \(\mu\) as defined in Eq.~(\ref{12}). Both
quantities are exact at leading order in \(\alpha\), and the second
one scales as \(M_{\rm remnant} \propto \alpha^{(D-3)/2}\).

The ratio \(\alpha/M^2\) is not dimensionless once \(D \neq 4\). In
\(D\) spacetime dimensions the mass carries \([M] = L^{D-3}\) while
\([\alpha] = L^2\), so
\begin{equation}\label{eq:dimcount}
\left[\frac{\alpha}{M^2}\right] = L^{8-2D},
\end{equation}
which is dimensionless only for \(D=4\). Comparing the raw ratio
across dimensions therefore compares quantities of different physical
dimension, and the ``dichotomy'' that such a comparison produces is an
artifact of the units. Restoring the missing powers with the one
length scale the problem supplies, the remnant radius itself, we
define
\begin{equation}\label{eq:Rtilde}
\tilde{\mathcal{R}} \equiv \frac{\alpha}{M_{\rm remnant}^2}\,
r_0^{\,2D-8}.
\end{equation}
Substituting Eq.~\eqref{eq:r0exact} into Eq.~\eqref{eq:Rtilde}, every
power of \(\alpha\) cancels and we are left with
\begin{equation}\label{eq:Rtildeclosed}
\tilde{\mathcal{R}} = \frac{4\mu^2 d^3 (d-2)}{(2d-2)^4},
\end{equation}
a pure number in each dimension. Table~\ref{tab:wgcscale} collects
the results.

\begin{table}[h]
\centering
\caption{Remnant radius, remnant mass, the un-scaled ratio
\(\mathcal{R} = \alpha/M_{\rm remnant}^2\) and the dimensionless
combination \(\tilde{\mathcal{R}}\) of Eq.~\eqref{eq:Rtilde}, obtained
from the degeneracy conditions \eqref{eq:degen} with the untruncated
metric function. The leading behavior is independent of the sign and
magnitude of \(\Lambda\).}
\label{tab:wgcscale}
\begin{tabular}{c c c c c}
\hline
\(D\) & \(r_0^2\) & \(M_{\rm remnant}\) & \(\mathcal{R} = \alpha/M_{\rm remnant}^2\) & \(\tilde{\mathcal{R}}\) \\
\hline
\(4\) & \(\dfrac{16\alpha}{3}\) & \(\dfrac{8\sqrt{3}}{9}\sqrt{\alpha}\) & \(\dfrac{27}{64}\) & \(\dfrac{27}{64} \simeq 0.4219\) \\[4pt]
\(5\) & \(\dfrac{9\alpha}{2}\) & \(\dfrac{81\pi}{32}\alpha\) & \(\dfrac{1024}{6561\pi^2\alpha}\) & \(\dfrac{512}{729\pi^2} \simeq 0.0712\) \\[4pt]
\(6\) & \(\dfrac{64\alpha}{15}\) & \(\dfrac{8192\sqrt{15}\,\pi}{3375}\alpha^{3/2}\) & \(\dfrac{759375}{67108864\pi^2\alpha^2}\) & \(\dfrac{3375}{16384\pi^2} \simeq 0.0209\) \\[4pt]
\(7\) & \(\dfrac{25\alpha}{6}\) & \(\dfrac{15625\pi^2}{1728}\alpha^{2}\) & \(\dfrac{2985984}{244140625\pi^4\alpha^3}\) & \(\dfrac{13824}{15625\pi^4} \simeq 0.0091\) \\
\hline
\end{tabular}
\end{table}

Read as a whole, the table says something simple. The un-scaled ratio
is constant only in four dimensions and grows as \(\alpha^{4-D}\)
elsewhere, but that growth carries the dimension \(L^{8-2D}\) and is
not a physical statement. Once the ratio is made dimensionless the
apparent split disappears: \(\tilde{\mathcal{R}}\) settles on a
finite, \(\alpha\)-independent number in every case, so the effective
charge-to-mass ratio of the endpoint is fixed by the dimension alone.
The numbers decrease with \(D\), from \(27/64\) in four dimensions to
roughly \(0.009\) in seven, which means the effective bound weakens as
the spacetime opens up. Four dimensions are not qualitatively
special here, but they are quantitatively the most constrained.

A second feature follows from Eq.~\eqref{eq:r0exact} without further
work. Since \(r_0\) and \(M_{\rm remnant}\) both vanish with
\(\alpha\), the endpoint is a purely quantum object in every
dimension; switching off the area gap removes it entirely rather than
leaving a classical relic. The cosmological constant plays no role in
this statement, which is worth stressing because the numerical work
of Secs.~\ref{isec5} uses \(\Lambda < 0\) throughout.

The same data also fix the response of the mass to a change in the
quantum parameter. Differentiating the exact mass function at fixed
radius and evaluating it at \(r_0\), which we carry out in
Appendix~B, gives the compact identity
\begin{equation}\label{eq:URM}
\alpha \left( \frac{\partial M}{\partial \alpha} \right)_{r=r_0}
= \frac{D-3}{2}\, M_{\rm remnant},
\qquad\text{equivalently}\qquad
\mathcal{R}\,\mathcal{U}\, M_{\rm remnant} = \frac{D-3}{2},
\end{equation}
where \(\mathcal{U}\) is the universal combination constructed in the
next section. Equation~\eqref{eq:URM} ties the thermodynamic response
of the remnant to the WGC-like ratio in a way that holds dimension by
dimension, with no free constants. We verify it numerically in
Table~\ref{tab:numU}.

To summarize, treating \(\alpha\) as an effective charge and the
zero-temperature endpoint as an effective extremal state produces a
scaling law that survives dimensional analysis and admits closed-form
coefficients. It is an analogy and not a derivation: there is no gauge
field, the remnant is macroscopic rather than a particle in a
spectrum, and nothing here constrains the light states that the
conjecture is actually about. What the analogy does supply is a sharp
consistency target. Any effective description that produces a
zero-temperature remnant from a minimal-area-gap correction should
reproduce Eq.~\eqref{eq:Rtildeclosed}, and a model that does not is
either using a different quantum correction or is not reaching the
same endpoint.

\section{Universal Relation for Quantum-Corrected Black Holes}\label{isec4}
The universal extremality relation connects entropy, temperature and
mass in the limit where the temperature vanishes. It has been checked
in Einstein gravity, in higher-derivative theories and in models with
nonlinear electrodynamics, and its appeal is that it follows from the
first law rather than from the details of any particular action. We
now ask what it becomes for the quantum-corrected black hole
constructed above, where the extremal limit is supplied by the
remnant.

We deform the theory by shifting the quantum parameter,
\(\alpha \to \alpha + \varepsilon\), with \(\varepsilon\) small, and
we track the combination \(-T (\partial S/\partial \varepsilon)_M\)
as the remnant is approached. Writing the perturbed metric function
as
\begin{equation}
f(r,\varepsilon) = f_0(r) + \varepsilon \, \delta f(r) +
\mathcal{O}(\varepsilon^2),
\end{equation}
with \(f_0\) the unperturbed function of Sec.~\ref{isec2}, the
horizon radius shifts as well. Expanding \(f(r_h(\varepsilon),
\varepsilon) = 0\) to first order and using \(f_0(r_h(0)) = 0\),
\begin{equation}
\frac{dr_h}{d\varepsilon} = - \frac{ \left( \partial f / \partial
\varepsilon \right)_{r_h(0)} }{ \left( \partial f_0 / \partial r
\right)_{r_h(0)} }.
\end{equation}
The denominator is \(2\kappa_0 = 4\pi T_0\), which vanishes at the
remnant. The horizon displacement is therefore not analytic in
\(\varepsilon\) there, and the limit has to be taken on a combination
in which the vanishing denominator cancels. That combination is the
one we compute.

From the first law in the extended phase space,
\begin{equation}
dM = T dS + V dP,
\end{equation}
and holding \(\Lambda\) fixed so that \(dP = 0\), the standard
manipulation with \(M = M(S,\varepsilon)\) gives
\begin{equation}
\left( \frac{\partial S}{\partial \varepsilon} \right)_M
= - \frac{1}{T} \left( \frac{\partial M}{\partial \varepsilon}
\right)_S ,
\end{equation}
so that
\begin{equation}\label{eq:Udef}
\mathcal{U} \equiv \lim_{M \to M_{\rm remnant}} \left[ -T \left(
\frac{\partial S}{\partial \varepsilon} \right)_{M} \right] = \left(
\frac{\partial M}{\partial \varepsilon} \right)_S .
\end{equation}
The factor of \(T\) cancels the divergence, which is what makes
\(\mathcal{U}\) finite at an extremal point. This step uses only the
first law and is independent of the theory.

Evaluating \(( \partial M / \partial \varepsilon )_S\) requires care,
because the entropy of Table~\ref{tab:entropy} depends on \(\alpha\)
as well as on \(r_h\). Holding \(S\) fixed while varying \(\alpha\)
therefore moves the horizon, and in general
\begin{equation}\label{eq:chain}
\left( \frac{\partial M}{\partial \alpha} \right)_S
= \left( \frac{\partial M}{\partial \alpha} \right)_r
- \left( \frac{\partial M}{\partial r} \right)_\alpha
\frac{(\partial S/\partial \alpha)_r}{(\partial S/\partial
r)_\alpha}.
\end{equation}
At the remnant the second term switches off identically, since
\(T = 0\) is equivalent to \((\partial M/\partial r)_\alpha = 0\).
This is exact, not a leading-order statement, and it is the reason the
calculation can be carried out at fixed radius:
\begin{equation}\label{eq:Ualpha}
\mathcal{U} = \left( \frac{\partial M}{\partial \alpha}
\right)_{r=r_0} .
\end{equation}

For the explicit evaluation we again use the untruncated metric
function rather than the first-order mass of Table~\ref{tab:mass},
for the reason given in Sec.~\ref{isec3}. Solving \(f(r)=0\) for the
mass in the regime where the cosmological-constant terms are
subleading gives the closed form
\begin{equation}\label{eq:Mexact}
M(r,\alpha) = \frac{r^{d} - r^{d-1}\sqrt{r^2 - 4\alpha}}{4\mu\,
\alpha},
\end{equation}
which reduces to \(2 M \mu = r^{d-2}\) as \(\alpha \to 0\), as it
should. Differentiating Eq.~\eqref{eq:Mexact} with respect to
\(\alpha\) and evaluating at the remnant radius of
Eq.~\eqref{eq:r0exact} yields the results collected in
Table~\ref{tab:Uvalues}; the intermediate steps are in Appendix~B.

\begin{table}[h]
\centering
\caption{The universal combination \(\mathcal{U} = (\partial M /
\partial \alpha)_{r=r_0}\) evaluated at the remnant, and the exact
identity relating it to the remnant mass. Only the five-dimensional
case is independent of \(\alpha\), and therefore of the deformation
parameter \(\varepsilon\).}
\label{tab:Uvalues}
\begin{tabular}{c c c}
\hline
\(D\) & \(\mathcal{U}\) & \(\alpha\,\mathcal{U}/M_{\rm remnant}\) \\
\hline
\(4\) & \(\dfrac{4\sqrt{3}}{9\sqrt{\alpha}}\) & \(\dfrac{1}{2}\) \\[4pt]
\(5\) & \(\dfrac{81\pi}{32}\) & \(1\) \\[4pt]
\(6\) & \(\dfrac{4096\sqrt{15}\,\pi}{1125}\sqrt{\alpha}\) & \(\dfrac{3}{2}\) \\[4pt]
\(7\) & \(\dfrac{15625\pi^2}{864}\,\alpha\) & \(2\) \\
\hline
\end{tabular}
\end{table}

The last column is the content of Eq.~\eqref{eq:URM}: in every
dimension \(\alpha \mathcal{U} = (D-3) M_{\rm remnant}/2\). We did
not expect the coefficient to come out this cleanly, and we have no
argument that it must; it is an observation about this family of
solutions, checked case by case.

Two consequences follow. First, \(\mathcal{U}\) is finite at the
remnant in every dimension, so the universal extremality relation
does survive the loop-quantum-gravity correction. Second, and less
comfortably, \(\mathcal{U}\) is not independent of the deformation.
Only for \(D=5\) does the \(\alpha\)-dependence cancel; elsewhere
\(\mathcal{U}\) inherits it through \(r_0(\alpha)\), and since
\(\alpha \to \alpha + \varepsilon\), it varies with \(\varepsilon\).
Table~\ref{tab:numU} shows the effect in four dimensions with
\(\Lambda = -0.01\) and \(\alpha_0 = 0.01\), where the remnant is
located by solving Eq.~\eqref{eq:degen} numerically with all
\(\Lambda\) terms retained.

\begin{table}[h]
\centering
\caption{Four-dimensional remnant data as a function of the
deformation \(\varepsilon\), for \(\Lambda = -0.01\) and \(\alpha_0 =
0.01\). The last column tests the identity \(\mathcal{R}\,
\mathcal{U}\, M_{\rm remnant} = (D-3)/2 = 1/2\); the residual excess
of a few parts in \(10^4\) is the \(\Lambda\)-dependent correction
neglected in Eq.~\eqref{eq:r0exact}.}
\label{tab:numU}
\begin{tabular}{c c c c c c}
\hline
\(\varepsilon\) & \(\alpha = \alpha_0 + \varepsilon\) & \(r_0\) & \(M_{\rm remnant}\) & \(\mathcal{U}\) & \(\mathcal{R}\,\mathcal{U}\,M_{\rm remnant}\) \\
\hline
\(0.001\) & \(0.011\) & \(0.2422\) & \(0.1615\) & \(7.3430\) & \(0.5001\) \\
\(0.002\) & \(0.012\) & \(0.2530\) & \(0.1687\) & \(7.0307\) & \(0.5002\) \\
\(0.005\) & \(0.015\) & \(0.2828\) & \(0.1886\) & \(6.2892\) & \(0.5002\) \\
\(0.010\) & \(0.020\) & \(0.3266\) & \(0.2178\) & \(5.4477\) & \(0.5003\) \\
\hline
\end{tabular}
\end{table}

Over the range shown, \(\mathcal{U}\) falls from \(7.34\) to
\(5.45\), a change of about \(26\%\), while the combination
\(\mathcal{R} \mathcal{U} M_{\rm remnant}\) stays at \(1/2\) to four
significant figures. The reading we take from this is that the
invariant object is not \(\mathcal{U}\) on its own but the product in
Eq.~\eqref{eq:URM}. Universality survives the deformation; it simply
lives in a different combination than one might first guess, and
checking it requires tracking \(r_0(\alpha)\) rather than treating
the remnant radius as fixed background data.

\section{Bondi Accretion}\label{isec5}
\subsection{Theoretical Framework of Accretion Fluid}
We consider two distinct fluid models: the cosmological dark fluid
and the exponential density profile. The dark fluid model is of
particular cosmological interest because it represents a significant
departure from the standard $\Lambda$CDM paradigm and has been
investigated in connection with scalar field theories, inflationary
dynamics, and the cosmological constant problem \cite{1,2,3,4,5,6}.
In contrast, the exponential density profile was originally
introduced in \cite{7} and further developed by Sofue \cite{8,9} to
explain the observed rotation curves of galaxies. Owing to its
astrophysical relevance, this profile has also been widely employed
in studies of accretion disc dynamics and related gravitational
phenomena \cite{10}. Accretion around black holes involves
specifying the energy-momentum tensor. If the matter part is assumed
as perfect fluid, then the energy-momentum tensor becomes
\begin{eqnarray}\label{1p}
T_{\mu\nu}=(\rho+P)u^{\mu}u^{\nu}-Pg_{\mu\nu},
\end{eqnarray}
the quantities $\rho$, $P$, and $u^{\mu}$ denote the density,
pressure, and four-velocity of the accreting fluid, respectively.
The four-velocity takes the general form
\begin{eqnarray}\label{2p}
u^{\mu}=\frac{dx^\mu}{d\tau}=(u^t,u^r,0,0),
\end{eqnarray}
here, $\tau$ denotes the proper time, with $u^\theta=u^\phi=0$.
Imposing the normalization condition for the four-velocity, $u^\mu
u_\mu=-1$, yields
\begin{eqnarray}\label{3p}
-f(r)(u^t)^2+\frac{(u^r)^2}{g(r)}=-1,
\end{eqnarray}
hence, the time component of the four-velocity can be expressed as
\begin{eqnarray}\label{4p}
u^t=\pm\sqrt{\frac{(u^r)^2+g(r)}{f(r)g(r)}},
\end{eqnarray}
for simplicity, we set $u^r=u$. The square-root term admits two
possible branches, corresponding to $(u^t>0)$ and $(u^t<0)$, which
represent forward and backward evolution in time, respectively. To
preserve causality, only the positive branch $(u^t>0)$ is physically
admissible, ensuring that the fluid falls into the black hole rather
than escaping from it. Moreover, the condition $(u<0)$ characterizes
inward accretion, whereas $(u>0)$ corresponds to an outward fluid
flow. To describe relativistic fluid accretion, we restrict our
analysis to the equatorial plane by setting $\theta=\pi/2$, for
which the metric determinant simplifies to
$\sqrt{-g}=r^2\sqrt{f(r)g(r)^{-1}}$. We then apply the conservation
law of the energy--momentum tensor,
$T^{\mu\nu}_{;\mu}=\frac{1}{\sqrt{-g}}\partial_{r}(\sqrt{-g}T^{\mu\nu})$,
we obtain
\begin{eqnarray}\label{5p}
(P+\rho) u r^2 \frac{f(r)}{g(r)}\sqrt{(u)^2+g(r)}=g_0,
\end{eqnarray}
here, $g_0$ is an integration constant. Contracting the conservation
equation onto the four-velocity vector, $u_\mu
T^{\mu\nu}_{;\mu}=u^\mu\partial_{\mu}\rho+(P+\rho)\nabla_{\mu}u^\mu=0$,
reduces to simplified form
\begin{eqnarray}\label{6p}
\frac{\rho^\prime}{P+\rho}+\frac{u^\prime}{u}+\frac{f^{\prime}(r)}{2f(r)}+\frac{g^\prime(r)}{2g(r)}+\frac{2}{r}=0.
\end{eqnarray}
The prime notation indicates differentiation with respect to r. Eq.
(\ref{6p}) can then be integrated to obtain
\begin{eqnarray}\label{7p}
 u r^2 \sqrt{\frac{E(r)}{F(r)}}e^{\big(\int\frac{d\rho}{P+\rho}\big)}=-g_1,
\end{eqnarray}
the constant $g_1$ arises as an integration constant. Imposing the
inward flow condition $(u<0)$ requires $g_1$ to be negative. The
mathematical expression governing the mass flux
$J^{\mu}_{;\mu}=\frac{1}{\sqrt{-g}}\frac{d}{dr}(J^r\sqrt{-g})=0$,
carrying out the integration yields
\begin{eqnarray}\label{8p}
 \rho u r^2 \sqrt{\frac{f(r)}{g(r)}}=g_2,
\end{eqnarray}
here, $g_2$ denotes an integration constant. Combining Eqs.
(\ref{5p}) and (\ref{7p}) gives
\begin{eqnarray}\label{9p}
(P+\rho)\sqrt{(u)^2+g(r)}\sqrt{\frac{f(r)}{g(r)}}e^{\big(-\int\frac{d\rho}{P+\rho}\big)}=g_3,
\end{eqnarray}
the constant $g_3$ is expressed in terms of $g_0$ and $g_1$. Taking
the ratio of Eqs. (\ref{5p}) and (\ref{8p}) yields
\begin{eqnarray}\label{10p}
\frac{(P+\rho)}{\rho}\sqrt{(u)^2+g(r)}\sqrt{\frac{f(r)}{g(r)}}=g_4.
\end{eqnarray}
A new arbitrary constant, $g_4$ , is introduced, which is determined
by $g_0$ and $g_2$.

\subsection{Dark Fluid Equation of State}

To formulate the governing equations of the dark fluid accretion
process, we consider the following equation of state (EoS) for the
pressure, which relates the fluid pressure to its energy density and
forms the basis of the subsequent derivation
\begin{eqnarray}\label{11p}
P=\omega\rho=constant,
\end{eqnarray}
The parameter $\omega$ represents the constant barotropic parameter
\cite{11}. As the first case, we consider a fluid characterized by
the following equation of state (EoS). The associated expressions
for the radial velocity, energy density, and mass accretion rate are
obtained accordingly. In particular, combining Eqs. (\ref{7p}) and
(\ref{11p}) gives the following expression for the radial velocity
\cite{12}
\begin{eqnarray}\label{12p}
u=\pm\sqrt{\frac{g(r)}{f(r)}g_3^{2}-g(r)}.
\end{eqnarray}
The negative sign is chosen to describe the inward accretion flow.
Using Eq. (\ref{8p}) to evaluate the energy density $\rho$ and
substituting the radial velocity from Eq. (\ref{12p}) leads to the
following expression \cite{12}.
\begin{eqnarray}\label{13p}
\rho=-\frac{g_2}{r^2}\frac{\sqrt{\frac{g(r)}{f(r)}}}{\sqrt{\frac{g(r)}{f(r)}g_3^2-g(r)}}.
\end{eqnarray}
The Bondi accretion rate is obtained by integrating the fluid energy
flux across the two-dimensional event horizon of the black hole
\cite{13}. The rate of change of the black hole mass is expressed as
$\dot{M}=-\int T^{1}_{0}\sqrt{-g} d\theta d\phi$. The following
general expressions are derived in the context of dark fluid
accretion \cite{12}
\begin{eqnarray}\label{14p}
\dot{M}=-4\pi r^2(P+\rho)\frac{u\sqrt{u^2+g(r)}}{f(r)g(r)}.
\end{eqnarray}
Using Eq. (\ref{14p}), the rate of change of the black hole mass can
be expressed as follows
\begin{eqnarray}\label{15p}
\dot{M}=4\pi g_1 M^2(\rho_\infty+P_\infty),
\end{eqnarray}
the derived result holds for any equation of state defined by
$P=P(\rho)$. Therefore, the rate of change of the black hole mass
due to fluid accretion is given by
\begin{eqnarray}\label{16p}
\dot{M}=4\pi g_1 M^2(\rho+P).
\end{eqnarray}
The sign of $\rho+P$ determines the evolution of the black hole mass
during the accretion process. For fluids satisfying the condition
$\rho+P>0$, the accretion increases the black hole mass. In
contrast, the accretion of phantom dark energy, characterized by
$\rho+P<0$, leads to a continuous decrease in the black hole mass.

\subsection{Exponential Density Profile}

The exponential density profile, originally introduced in Refs.
\cite{8, 9}, provides a physically motivated description of the dark
matter distribution surrounding the black hole. In this model, the
matter density decreases exponentially with the radial distance from
the center and is expressed as
\begin{eqnarray}\label{17p}
\rho(r)=\rho_0 e^{-\frac{r}{r_0}},
\end{eqnarray}
The parameters $r_0$ and $\rho$ represent the core radius and the
characteristic density, respectively. The following analysis
considers the second fluid, characterized by the exponential density
profile. By combining Eqs. (\ref{5p}) and (\ref{8p}), the
corresponding expressions for the radial velocity and pressure are
obtained \cite{12}
\begin{eqnarray}\label{18p}
u=\frac{g_2}{\rho u r^2}\sqrt{\frac{f(r)}{g(r)}},
\end{eqnarray}
\begin{eqnarray}\label{19p}
P=\bigg(\frac{g_2/g_0}{\sqrt{g_2^2/\rho r^4+f(r)}}-1\bigg)\rho.
\end{eqnarray}
To describe the present scenario, Eq. (\ref{15p}) is extended to a
general equation of state satisfying $P=P(\rho)$, leading to the
following expression
\begin{eqnarray}\label{20p}
\dot{M}_{exponential}=4\pi g_1 M^2(P(r)+\rho(r)).
\end{eqnarray}

\subsection{Accretion Disk Luminosity}

The luminosity generated by the accretion process is evaluated using
the Eddington luminosity expression, which provides the theoretical
upper bound on the radiative output of an accreting black hole. This
limiting luminosity corresponds to the condition under which the
outward radiation pressure acting on the infalling matter exactly
balances the inward gravitational attraction. As a result, the
accretion flow remains stable only below this threshold, whereas
luminosities exceeding the Eddington limit produce sufficient
radiation pressure to oppose gravity and significantly suppress or
even terminate the accretion process \cite{14}. The Eddington
luminosity is obtained by balancing the outward radiation pressure
exerted by the luminous source against the inward gravitational
attraction acting on a proton. Because photons interact primarily
with electrons through Thomson scattering, the strong electrostatic
coupling between electrons and protons causes the entire plasma to
experience the same radiative force. The resulting equilibrium
defines the Eddington limit and yields the following expression
\begin{eqnarray}\label{21p}
L_{Edd}=\frac{4\pi GMm_p}{\sigma_T},
\end{eqnarray}
where $G, m_p$ and $\sigma_T$ represent the gravitational constant,
the proton mass and the Thomson scattering cross section,
respectively. The radiative efficiency of the accretion process is
characterized by the efficiency parameter $\eta_{eff}$, commonly
assumed to have a value of approximately 0.1, which quantifies the
fraction of the accreted rest-mass energy converted into radiation.
Accordingly, the final expression for the luminosity is given by
\begin{eqnarray}\label{22p}
L=\eta_{eff} \dot{M}.
\end{eqnarray}
The quantity $\dot{M}$ corresponds to the mass accretion rate
obtained from Eq. (\ref{14p}) for the first case and Eq. (\ref{20p})
for the second case.

\subsection{Critical Points Analysis and Variable Expression}

The analysis now turns to the determination of the critical points,
which correspond to the radii where the radial velocity of the
accreting fluid becomes identical to the local sound speed, marking
the transonic transition of the flow. As the matter is continuously
attracted by the gravitational field of the black hole, its inward
radial velocity progressively increases while propagating along the
streamlines toward the event horizon. At a particular location, the
fluid undergoes a smooth transition from the subsonic regime to the
supersonic regime, and this radius is identified as the critical
point governing the overall accretion dynamics. Therefore,
investigating these critical conditions clarifies the behavior and evolution of the accreting matter as it
responds to the strong gravitational influence in the vicinity of
the black hole. We investigate the critical behavior of the
accretion flow by adopting the conventional approach developed for
spherically symmetric black hole accretion \cite{15}, where the
fluid dynamics are determined through the fundamental conservation
equations governing mass and energy transport. Taking the
logarithmic derivatives of Eqs. (\ref{8p})-(\ref{10p}) and
subsequently combining the resulting differential relations yields
the equation that determines the location of the critical radius,
$r_c$ as
\begin{eqnarray}\nonumber
\bigg[V^2&-&\frac{u^2}{u^2+g(r)}\bigg]\frac{du}{u}+\bigg[(V^2-1)\bigg(\frac{f^\prime(r)}{2f(r)}-\frac{g^\prime(r)}{2g(r)}\bigg)\\\label{23p}&+&\frac{2}{r}V^2
-\frac{g^\prime(r)}{2(u^2+g(r))}\bigg]dr=0,
\end{eqnarray}
where $V^2=\frac{d ln(\rho+P)}{d ln\rho}-1$. The location of the
critical point is established by requiring both bracketed terms to
vanish simultaneously. Subsequently, isolating the variables $u^2$
and $V^2$ decouples the governing equations, yielding a closed
system of independent relations that specifies the critical
properties of the accreting fluid
\begin{eqnarray}\label{24p}
V_c^2=\frac{1}{1+\frac{4f(r_c)}{f^{\prime}(r_c) r_c}},\
u_c^2=\frac{g(r_c)f^{\prime}(r_c) r_c}{4f(r_c)}.
\end{eqnarray}
For the special case in which $f(r)=g(r)$, the spacetime metric
simplifies considerably, leading to a reduced form of the velocity
expression. By substituting this condition into Eq. (\ref{12p}), the
radial velocity can be written as follows \cite{12}
\begin{eqnarray}\label{25p}
u=-\sqrt{g_3^2-f(r)},
\end{eqnarray}
under the same assumption, Eq. (\ref{13p}) reduces to the following
expression for the energy density \cite{12}
\begin{eqnarray}\label{26p}
\rho=-\frac{g_2}{r^2\sqrt{g_3^2-f(r)}},
\end{eqnarray}
similarly, the mass accretion rate derived from Eq. (\ref{14p})
which is expressed as follows \cite{12}
\begin{eqnarray}\label{27p}
\dot{M}=\frac{4\pi r^2
g_3}{f(r)^2}\bigg(P\sqrt{g_3^2-f(r)}-\frac{g_2}{r^2}\bigg).
\end{eqnarray}
\subsection{Graphical Behavior}

In this section, we present a detailed graphical analysis of the
radial velocity, energy density, mass accretion rate, and luminosity
for the dark fluid model, together with the pressure, exponential
mass accretion rate, and exponential luminosity corresponding to the
exponential density model. The behavior of these physical quantities
is examined in higher-dimensional spacetimes by
varying the dimensional parameter over the range $D=3$ to $7$,
thereby illustrating the influence of spacetime dimensionality on
the accretion process.

\subsubsection{$d=3$ ($D=4$)}

By incorporating Eq. (\ref{1b11}) into Eqs. (\ref{22p}) and
(\ref{25p})-(\ref{27p}), we present the radial profiles of the fluid
velocity, energy density, mass accretion rate, and luminosity as
functions of the dimensionless radial coordinate r/M, as illustrated
in \textbf{Figures. 1 and 2}. The numerical analysis is performed by
fixing the model parameters to $M=1, ~g_2 =-5, ~g_3 =1.5, ~\eta=0.1,
~P=-0.4, ~G=1$ and $\Lambda=-0.1$. The radial velocity $u$ remains
negative across the entire range of $r/M$, demonstrating that the
fluid continuously undergoes inward accretion onto the black hole.
As the radial distance increases, the magnitude of the velocity
gradually decreases and approaches zero, indicating that the
infalling fluid slows down as it moves farther from the
gravitational source. An increase in the parameter $\alpha$ from
0.01 to 0.07 shifts the velocity profiles toward less negative
values, revealing that larger a weakens the radial infall at a given
location. In contrast, the energy density $\rho$ decreases steadily
with increasing $r/M$, attaining its maximum value close to the
event horizon where the gravitational influence is strongest and
progressively diminishing at larger distances. Although higher
values of $\alpha$ yield greater energy densities throughout the
flow, the distinction between the corresponding curves becomes less
significant in the outer region. Meanwhile, the mass accretion rate
$\dot{M}$ increases monotonically with the radial coordinate,
exhibiting a more pronounced growth near the upper limit of the
plotted domain, and its magnitude is consistently enhanced as a
increases, signifying a more efficient transfer of matter onto the
black hole. The luminosity $L$ follows an analogous behavior, rising
continuously with $r/M$ and becoming larger for increasing values of
$\alpha$. Since the emitted luminosity is directly proportional to
the mass accretion rate, the increase in $\dot{M}$ associated with
larger a naturally gives rise to a corresponding enhancement in the
radiative output of the accreting system. Moreover, inserting Eqs.
(\ref{1b11}, \ref{17p}) in (\ref{19p}, \ref{20p}, \ref{22p}), the
pressure distribution is evaluated by choosing $g_0 =1, ~g_1 =-2,
~r_0=2$ and $\rho_0=0.4$, the resulting profile is displayed in
\textbf{Figure 3}. The pressure remains negative over the entire
range of $r/M$, confirming the presence of a negative-pressure fluid
surrounding the black hole. As the radial distance increases, the
pressure decreases slightly to a minimum before gradually increasing
toward less negative values. Increasing the parameter $\alpha$
shifts the pressure profiles upward without noticeably altering
their overall shape, indicating that $\alpha$ primarily influences
the magnitude of the pressure rather than its radial evolution. In
contrast, both the exponential mass accretion rate, $\dot{M}_{exp}$,
and the corresponding exponential luminosity, $L_{exp}$, decrease
monotonically with increasing $r/M$, demonstrating that the
accretion efficiency and radiative output weaken at larger distances
from the black hole. Higher values of $\alpha$ lead to slightly
smaller values of both $\dot{M}_{exp}$ and $L_{exp}$, although the
differences among the curves remain modest, suggesting that the
influence of $\alpha$ on these quantities is relatively weak.
\begin{figure}
\epsfig{file=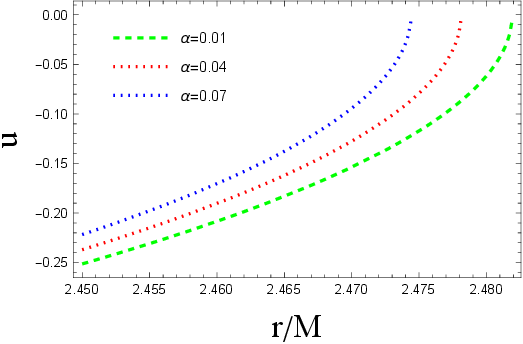, width=.45\linewidth, height=1.8in}
\epsfig{file=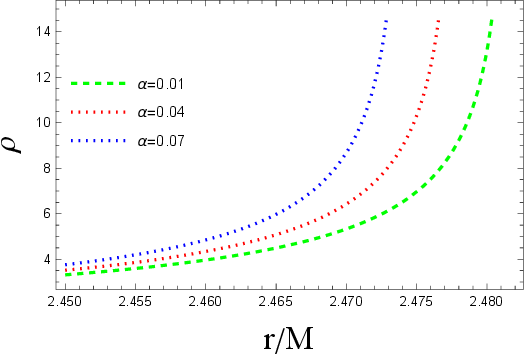, width=.45\linewidth, height=1.8in}
\caption{Radial velocity $u$ and energy density $\rho$ with respect
to $r/M$ for $d=3$ ($D=4$).} \epsfig{file=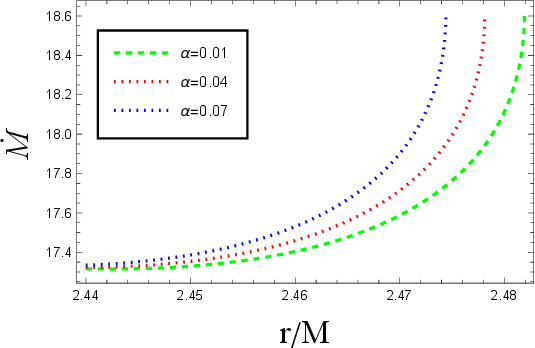, width=.45\linewidth,
height=1.8in} \epsfig{file=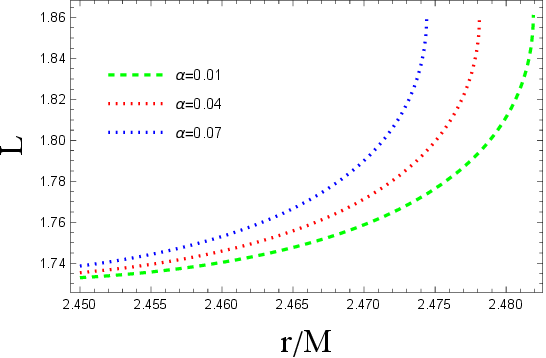, width=.45\linewidth,
height=1.8in} \caption{The mass accretion rate $\dot{M}$ and
luminosity $L$ with respect to $r/M$ for $d=3$ ($D=4$).}
\end{figure}
\begin{figure}
\epsfig{file=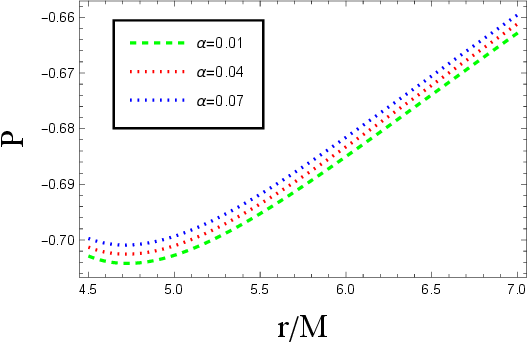, width=.3\linewidth, height=1.5in}
\epsfig{file=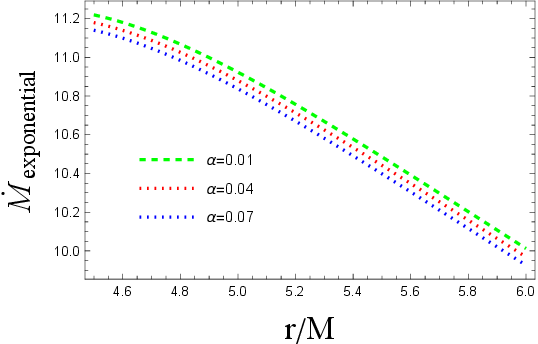, width=.3\linewidth, height=1.5in}
\epsfig{file=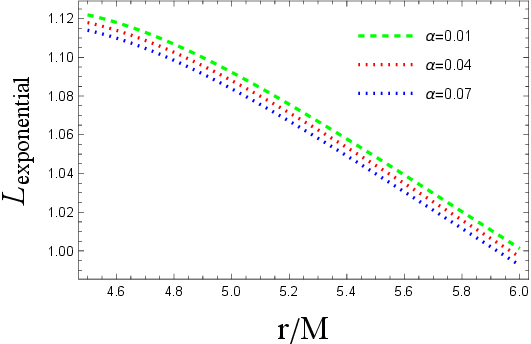, width=.3\linewidth, height=1.5in}
\caption{The pressure profile $P$, exponential mass rate
$\dot{M}_{exponential}$ and exponential luminosity $L_{exponential}$
with respect to $r/M$ for $d=3$ ($D=4$).}
\end{figure}

\subsubsection{$d=4$ ($D=5$)}

By substituting Eq. (\ref{1b11}) into Eqs. (\ref{22p}) and
(\ref{25p})-(\ref{27p}) and the radial profiles of the fluid
velocity, energy density, mass accretion rate, and luminosity are
obtained as functions of the dimensionless coordinate $r/M$, as
shown in \textbf{Figures 4 and 5}. For $D=4$, the radial velocity
remains negative throughout the domain, confirming inward accretion,
while its magnitude decreases from approximately -0.58 to -0.40 with
increasing $r/M$. The energy density decreases rapidly away from the
black hole, indicating a lower concentration of accreting matter at
larger radii. In contrast, $\dot{M}$ increases with increases in
$r/M$, with larger values of $\alpha$ producing a stronger accretion
rate. The luminosity $L$ exhibits a similar increasing trend,
reflecting its direct dependence on $\dot{M}$. Using Eqs.
(\ref{1b11}) and (\ref{17p}) in Eqs. (\ref{19p}), (\ref{20p}) and
(\ref{22p}), the pressure profile is obtained for $D=4$, as
displayed in \textbf{Figure 6}. The pressure remains negative
throughout the radial domain, characteristic of the negative
pressure fluid surrounding the black hole. Increasing $\alpha$
shifts the profiles upward while preserving their overall shape,
indicating a primarily magnitude dependent effect. For the
exponential case, both $\dot{M}_{exp}$ and $L_{exp}$ decrease
monotonically with $r/M$, while larger $\alpha$ values produce a
slight reduction in both quantities, demonstrating its relatively
weak influence on the exponential accretion and luminosity profiles.
\begin{figure}
\epsfig{file=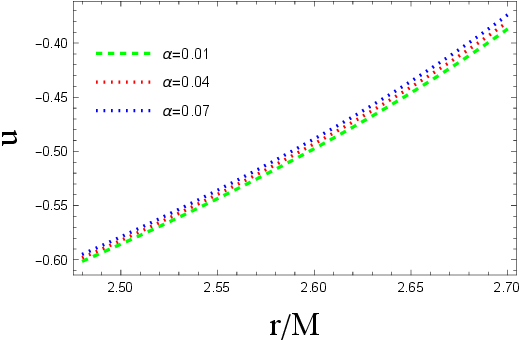, width=.45\linewidth, height=1.8in}
\epsfig{file=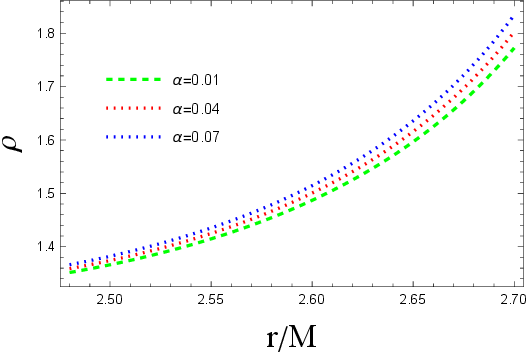, width=.45\linewidth, height=1.8in}
\caption{Radial velocity $u$ and energy density $\rho$ with respect
to $r/M$ for $d=4$ ($D=5$).} \epsfig{file=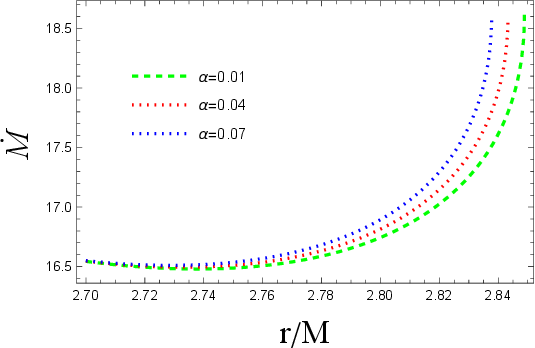, width=.45\linewidth,
height=1.8in} \epsfig{file=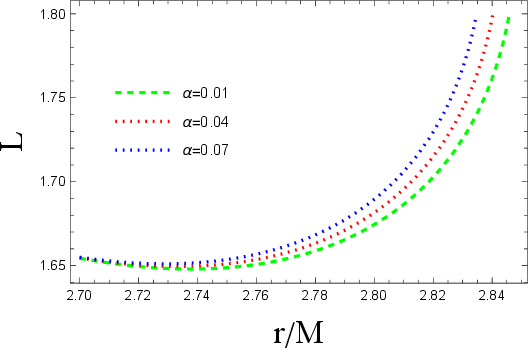, width=.45\linewidth,
height=1.8in} \caption{The mass accretion rate $\dot{M}$ and
luminosity $L$ with respect to $r/M$ for $d=4$ ($D=5$).}
\end{figure}
\begin{figure}
\epsfig{file=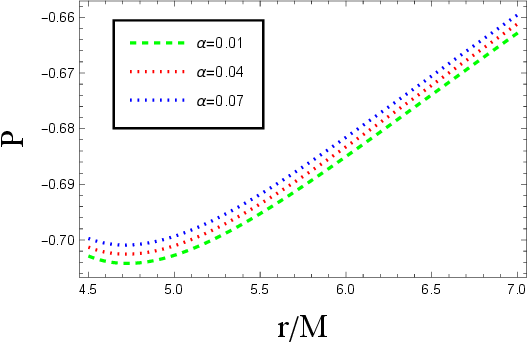, width=.3\linewidth, height=1.5in}
\epsfig{file=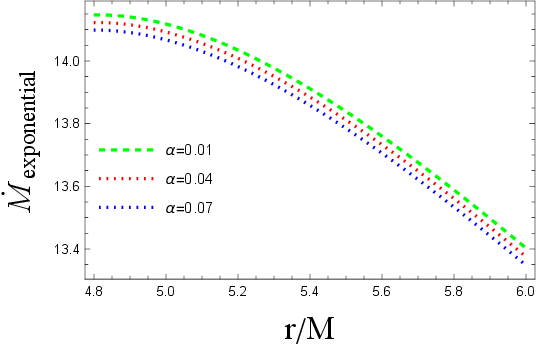, width=.3\linewidth, height=1.5in}
\epsfig{file=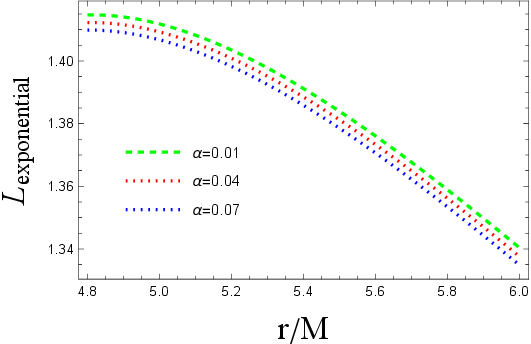, width=.3\linewidth, height=1.5in}
\caption{The pressure profile $P$, exponential mass rate
$\dot{M}_{exponential}$ and exponential luminosity $L_{exponential}$
with respect to $r/M$ for $d=4$ ($D=5$).}
\end{figure}

\subsubsection{$d=5$ ($D=6$)}

Substitution of Eq. (\ref{1b11}) into Eqs. (\ref{22p}) and
(\ref{25p})-(\ref{27p}) enables us to investigate the radial
evolution of the accreting fluid for $D=5$, with the corresponding
results presented in \textbf{Figures 7 and 8}. The radial velocity
$u$ remains negative throughout the investigated domain, confirming
the inward nature of the accretion flow. Its value changes from
nearly (-0.74) to (-0.70) as $r/M$ increases, indicating a gradual
reduction in the inflow speed away from the black hole. The energy
density exhibits an opposite trend, decreasing continuously with
radial distance. The mass accretion rate $\dot{M}$ grows with $r/M$,
with its increase becoming more evident toward the outer part of the
considered region. The luminosity $L$ follows the same qualitative
tendency as $\dot{M}$, increasing radially and attaining its largest
values for higher $\alpha$. This correlation reflects the direct
connection between the amount of accreted matter and the resulting
radiative output. The pressure profile is obtained by employing Eqs.
(\ref{1b11}) and (\ref{17p}) in Eqs. (\ref{19p}), (\ref{20p}) and
(\ref{22p}). The resulting behavior is depicted in \textbf{Figure
9}. Throughout the radial interval, $P$ retains negative values,
consistent with the presence of a negative-pressure fluid in the
vicinity of the black hole. The quantities $\dot{M}_{exp}$ and $L_{
exp}$ display a contrasting radial dependence: both decrease
steadily as the radial coordinate increases. Their maximum values
occur in the vicinity of the inner region, followed by a gradual
attenuation toward larger radii.
\begin{figure}
\epsfig{file=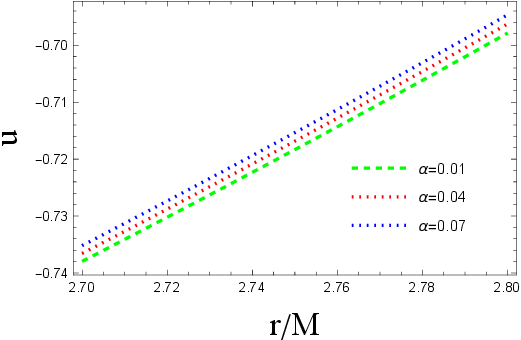, width=.45\linewidth, height=1.8in}
\epsfig{file=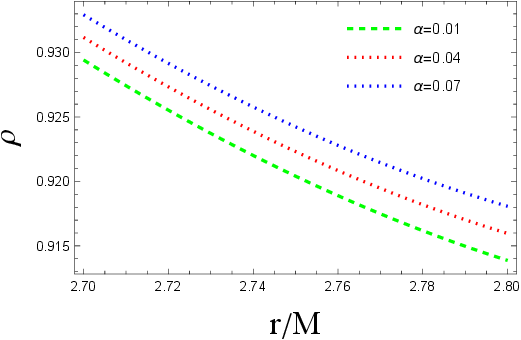, width=.45\linewidth, height=1.8in}
\caption{Radial velocity $u$ and energy density $\rho$ with respect
to $r/M$ for $d=5$ ($D=6$).} \epsfig{file=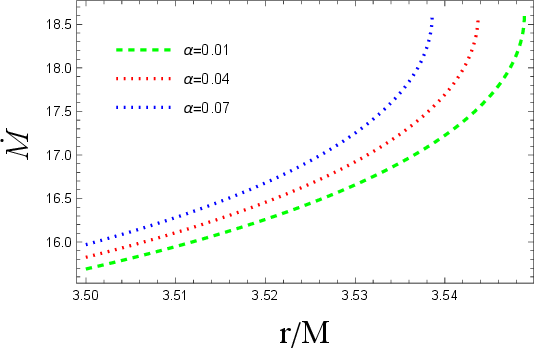, width=.45\linewidth,
height=1.8in} \epsfig{file=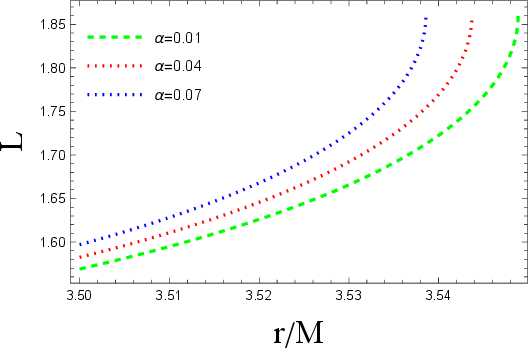, width=.45\linewidth,
height=1.8in} \caption{The mass accretion rate $\dot{M}$ and
luminosity $L$ with respect to $r/M$ for $d=5$ ($D=6$).}
\end{figure}
\begin{figure}
\epsfig{file=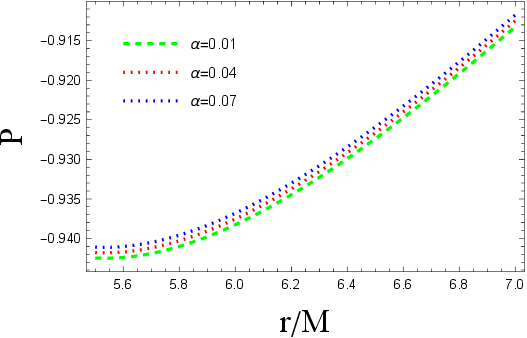, width=.3\linewidth, height=1.5in}
\epsfig{file=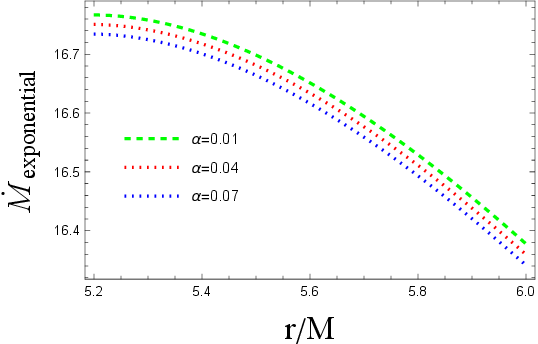, width=.3\linewidth, height=1.5in}
\epsfig{file=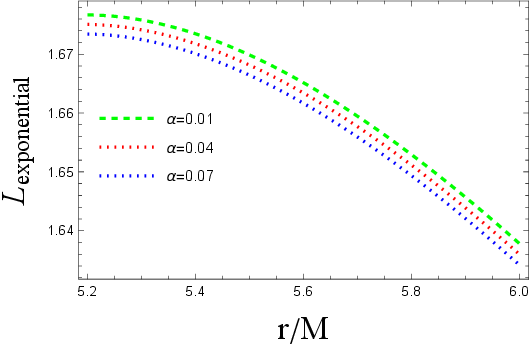, width=.3\linewidth, height=1.5in}
\caption{The pressure profile $P$, exponential mass rate
$\dot{M}_{exponential}$ and exponential luminosity $L_{exponential}$
with respect to $r/M$ for $d=5$ ($D=6$).}
\end{figure}

\subsubsection{$d=6$ ($D=7$)}

Putting Eq. (\ref{1b11}) into Eqs. (\ref{22p}) and
(\ref{25p})-(\ref{27p}) allows us to examine the radial behavior of
the accreting fluid for $D=6$, as illustrated in \textbf{Figures 10
and 11}. The radial velocity $u$ remains negative across the entire
domain, confirming inward accretion, while its magnitude decreases
as $r/M$ increases. In contrast, the energy density gradually
declines with radial distance. The mass accretion rate $\dot{M}$
increases outward, with a more pronounced rise in the outer region,
and the luminosity $L$ exhibits a similar trend, reaching higher
values for larger $\alpha$. Using Eqs. (\ref{1b11}) and (\ref{17p})
in Eqs. (\ref{19p}), (\ref{20p}) and (\ref{22p}), the pressure
profile is obtained and shown in \textbf{Figure 12}. The pressure
remains negative throughout the considered region, indicating the
presence of a negative-pressure fluid around the black hole. In
contrast, $\dot{M}_{exp}$ and $L_{exp}$ decrease monotonically with
increasing $r/M$, attaining their highest values near the inner
region and gradually diminishing at larger radii.
\begin{figure}
\epsfig{file=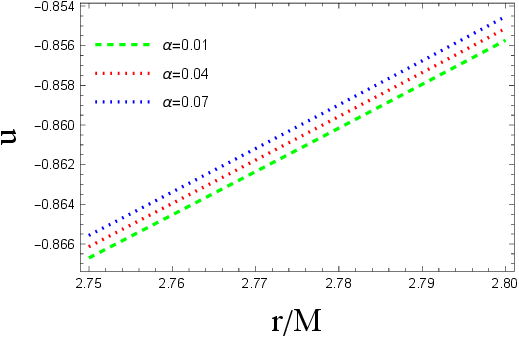, width=.45\linewidth, height=1.8in}
\epsfig{file=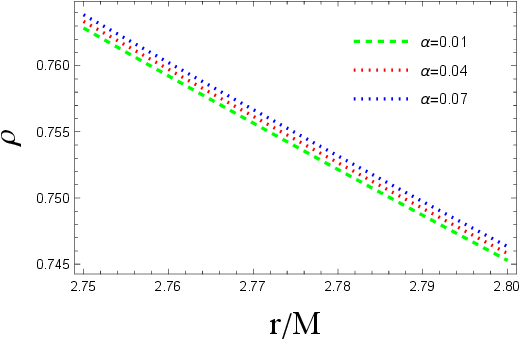, width=.45\linewidth, height=1.8in}
\caption{Radial velocity $u$ and energy density $\rho$ with respect
to $r/M$ for $d=6$ ($D=7$).} \epsfig{file=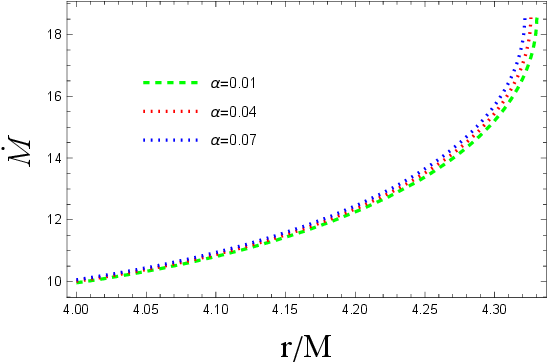, width=.45\linewidth,
height=1.8in} \epsfig{file=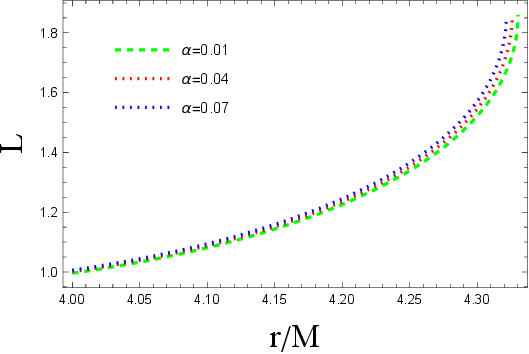, width=.45\linewidth,
height=1.8in} \caption{The mass accretion rate $\dot{M}$ and
luminosity $L$ with respect to $r/M$ for $d=6$ ($D=7$).}
\end{figure}
\begin{figure}
\epsfig{file=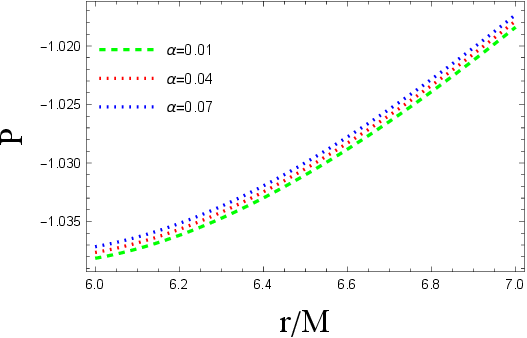, width=.3\linewidth, height=1.5in}
\epsfig{file=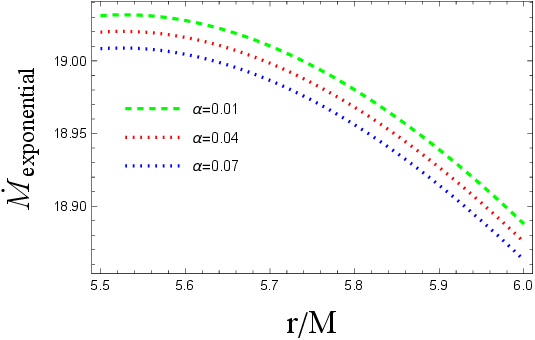, width=.3\linewidth, height=1.5in}
\epsfig{file=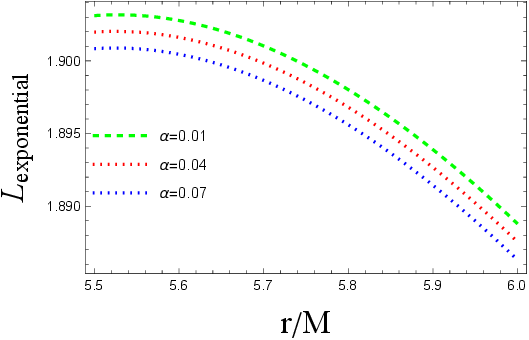, width=.3\linewidth, height=1.5in}
\caption{The pressure profile $P$, exponential mass rate
$\dot{M}_{exponential}$ and exponential luminosity $L_{exponential}$
with respect to $r/M$ for $d=6$ ($D=7$).}
\end{figure}

\subsubsection{$d=7$ ($D=8$)}

Using Eq. (\ref{1b11}) together with Eqs. (\ref{22p}) and
(\ref{25p})-(\ref{27p}), we obtain the radial characteristics of the
accreting fluid for $d=7$, as displayed in \textbf{Figures 13 and
14}. The velocity $u$ remains below zero over the entire radial
range, signifying a persistent inward flow. Meanwhile, the energy
density falls monotonically as the radial coordinate increases. The
accretion rate $\dot{M}$ shows an increasing profile, particularly
in the outer portion of the domain, and the luminosity $L$ mirrors
this behavior, with larger $\alpha$ producing greater radiative
output. The exponential-density case in this dimension follows the pattern
already established for $d=3$ through $d=6$: the pressure stays
negative across the sampled interval, while $\dot{M}_{exp}$ and
$L_{exp}$ fall off monotonically with $r/M$, so the sensitivity to
$\alpha$ remains weak.
\begin{figure}
\epsfig{file=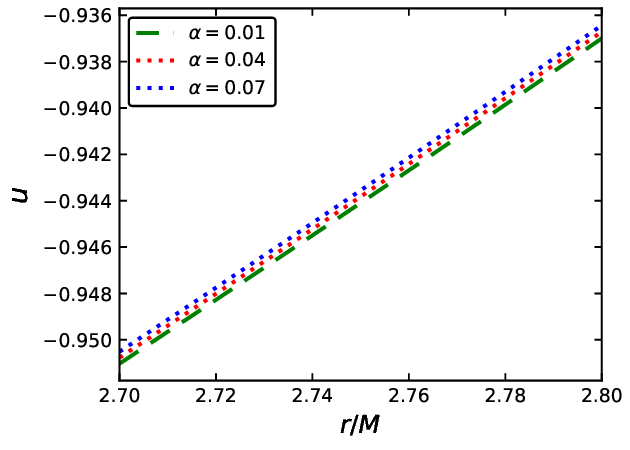, width=.45\linewidth, height=1.8in}
\epsfig{file=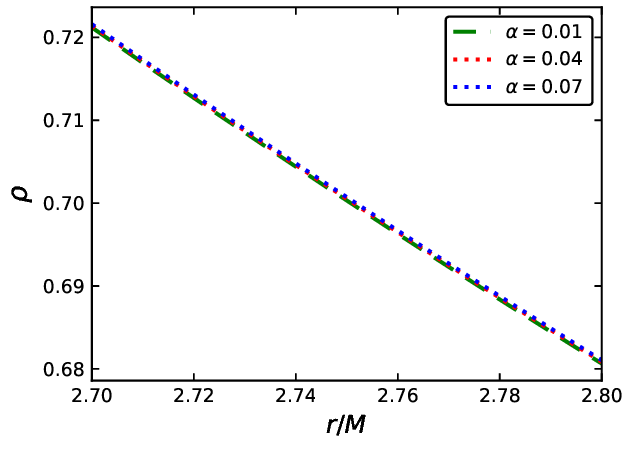, width=.45\linewidth, height=1.8in}
\caption{Radial velocity $u$ and energy density $\rho$ with respect
to $r/M$ for $d=7$ ($D=8$).} \epsfig{file=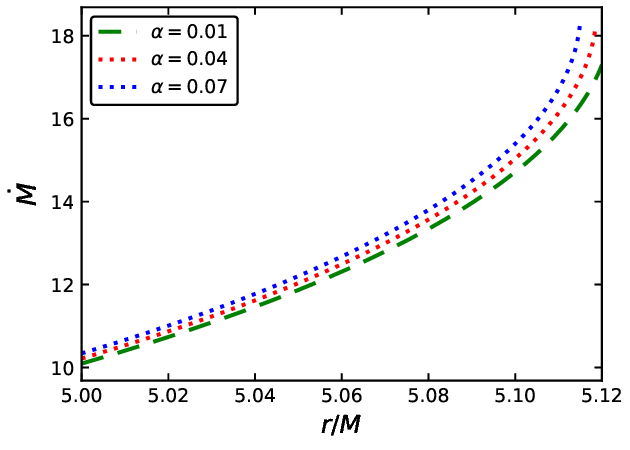, width=.45\linewidth,
height=1.8in} \epsfig{file=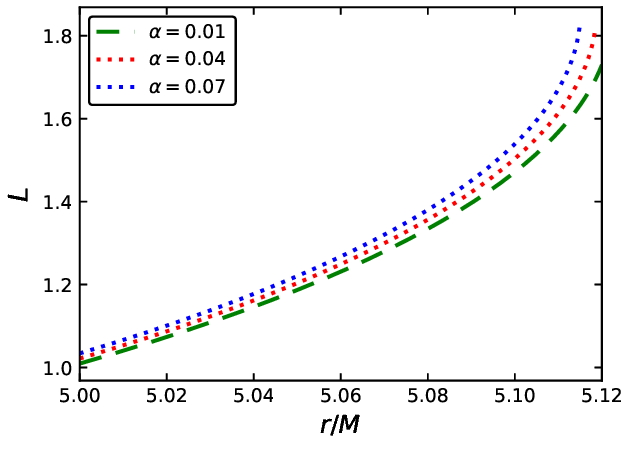, width=.45\linewidth,
height=1.8in} \caption{The mass accretion rate $\dot{M}$ and
luminosity $L$ with respect to $r/M$ for $d=7$ ($D=8$).}
\end{figure}

\subsection{Connection Between WGC-Like Behavior and Bondi Accretion}
\label{sec:connection}
We now connect the accretion formalism of this section to the
WGC-like relations of Sec.~\ref{isec3} by evaluating the flow at the
remnant radius \(r_0\), where the Hawking temperature vanishes. One
feature of that radius controls everything that follows. The remnant
is a degenerate horizon, so \(f(r_0) = 0\) identically once the mass
is set to \(M_{\rm remnant}\); the two conditions in
Eq.~\eqref{eq:degen} are not independent of the accretion problem but
sit right at its inner boundary. Any expression carrying an inverse
power of \(f\) therefore has to be handled as a limit rather than
evaluated naively at \(r_0\).

Two of the flow variables are already finite there. Using
Eq.~(\ref{25p}) with \(f(r_0) = 0\),
\begin{equation}\label{eq:u_r0_new}
u(r_0) = -\sqrt{g_3^2 - f(r_0)} = -g_3 ,
\end{equation}
so the infall speed at the remnant is fixed by the integration
constant alone and is independent of \(\alpha\), \(\Lambda\) and the
dimension. The density follows from Eq.~(\ref{26p}),
\begin{equation}\label{eq:rho_r0_new}
\rho(r_0) = -\frac{g_2}{r_0^2 \sqrt{g_3^2 - f(r_0)}}
= -\frac{g_2}{g_3\, r_0^2},
\end{equation}
which is finite and, through \(r_0\), carries the whole
\(\alpha\)-dependence of the near-remnant flow.

The mass accretion rate needs more care. The rate quoted in
Eq.~(\ref{27p}) carries an explicit factor \(f(r)^{-2}\) and so
appears to diverge at the remnant. That divergence is an artifact of
the normalization, not a physical statement about the flow. The
conserved quantity is the flux of the mixed component of the
energy-momentum tensor through a sphere of radius \(r\),
\begin{equation}\label{eq:Mdot_cons}
\dot{M} = -\Omega_{d-1} r^{d-1} T^{r}{}_{t}
= \Omega_{d-1} r^{d-1} (\rho + P)\, u\, u_t ,
\end{equation}
and for \(f = g\) one has \(u_t = -g_3\), while the continuity
equation \eqref{11p} fixes \(\rho u r^{d-1}\) to a constant. With the
barotropic equation of state \(P = \omega \rho\) this collapses to
\begin{equation}\label{eq:Mdot_r0_new}
\dot{M} = \Omega_{d-1}\, g_3\, g_2 \,(1+\omega),
\qquad
L = \eta_{\rm eff}\, \Omega_{d-1}\, g_3\, g_2\,(1+\omega),
\end{equation}
which is finite, radius-independent, and therefore takes the same
value at the remnant as anywhere else in the flow. This is the
expected behavior for a steady spherical accretion flow, where the
accretion rate is an eigenvalue of the problem rather than a function
of position, and it agrees with the form already quoted in
Eq.~(\ref{17p}). For \(d=3\) the geometric prefactor is
\(\Omega_2 = 4\pi\) and Eq.~\eqref{eq:Mdot_r0_new} reduces to
\(4\pi g_3 g_2 (1+\omega)\).

Putting Eq.~\eqref{eq:rho_r0_new} together with the remnant radius of
Eq.~\eqref{eq:r0exact} gives the relation we were after,
\begin{equation}\label{eq:rho_alpha}
\alpha\, \rho(r_0) = -\frac{g_2}{g_3}\,
\frac{d(d-2)}{(2d-2)^2}
= -\frac{g_2}{g_3}\, \frac{(D-1)(D-3)}{(2D-4)^2}.
\end{equation}
The product \(\alpha \rho(r_0)\) is a pure number times the
integration constants: it does not depend on \(\Lambda\), and its
dimensional dependence is weak, rising from \(3/16\) at \(D=4\) to
\(6/25\) at \(D=7\) and tending to \(1/4\) as \(D\) grows. The
physical content is direct. The density that the flow reaches at the
inner boundary is inversely proportional to the quantum parameter, so
a stronger area-gap correction produces a larger remnant and a more
dilute inner flow. Combining Eq.~\eqref{eq:rho_alpha} with
\(M_{\rm remnant} \propto \alpha^{(D-3)/2}\) gives
\(\rho(r_0) \propto M_{\rm remnant}^{-2/(D-3)}\), which is the
accretion counterpart of the WGC-like scaling of
Table~\ref{tab:wgcscale}.

Two limitations should be stated plainly. The flux integrals of
Sec.~\ref{isec5} were carried out on the equatorial slice with the
four-dimensional area element, and the Eddington luminosity
\eqref{22p} uses the four-dimensional Thomson opacity, so the
luminosities reported for \(d>3\) are four-dimensional radiative
estimates evaluated on a higher-dimensional background rather than
genuine \(D\)-dimensional Eddington limits. Restoring
\(\Omega_{d-1}\) and the \(D\)-dimensional opacity would change the
prefactors but not the \(\alpha\)-scaling, which is what
Eq.~\eqref{eq:rho_alpha} isolates. Second, the relations above hold at
the inner boundary of a steady flow; a genuinely time-dependent
accretion episode onto a near-remnant configuration is outside the
present framework.

\section{Conclusion and Results}
In this article we work out the effects of loop quantum gravity corrections on the gravitational collapse of a homogeneous dust sphere in higher dimensions. A quantum correction taken from the modified Friedmann equation of loop quantum cosmology is carried through an Oppenheimer-Snyder collapse, and the exterior geometry it leaves behind is static and controlled by the single deformation parameter \(\alpha = \gamma^2 \Delta^{2/(d-1)}\), with \(\gamma\) the Barbero-Immirzi parameter and \(\Delta\) the area gap. Closed-form expressions for the mass, the Hawking temperature and the entropy follow for \(3 \leq d \leq 6\), that is for spacetime dimensions four through seven. The temperature is not monotonic in the horizon radius: it starts from zero at a finite radius, rises to a peak, and then falls back onto the classical curve from above. Evaporation therefore halts at finite size, which is what makes the endpoint a remnant rather than a singularity. In the entropy the cosmological constant cancels exactly, leaving a dependence on \(r_h\) and \(\alpha\) alone, and the Bekenstein-Hawking area law is recovered as \(\alpha \to 0\). The four-dimensional case turns out to be the odd one: instead of a power-law correction it produces \(S = \pi r_h^2 + 4\pi \alpha \ln(r_h/r_\ast)\), a logarithmic term of the kind already familiar from independent microstate countings.

Treating \(\alpha\) as an effective charge and the zero-temperature endpoint as an effective extremal state then yields a scaling law of WGC type. Two things had to be handled carefully to get it right. First, the first-order mass expansion cannot be used at the remnant, because the quantum term in \(f(r)\) balances the classical one there rather than correcting it; we therefore impose \(f(r_0) = f'(r_0) = 0\) on the untruncated metric function, which gives \(r_0^2 = (2d-2)^2 \alpha / [d(d-2)]\) and \(M_{\rm remnant} \propto \alpha^{(D-3)/2}\) in closed form. Since \(r_0 \to 0\) with \(\alpha\), the cosmological-constant terms enter only at relative order \(\Lambda \alpha\), so the result holds for either sign of \(\Lambda\) and the earlier AdS-versus-dS question does not arise. Second, the ratio \(\alpha/M^2\) carries dimension \(L^{8-2D}\) and is dimensionless only at \(D=4\). Its apparent growth as \(\alpha^{4-D}\) in higher dimensions is a statement about units, not about physics. The dimensionless combination \(\tilde{\mathcal{R}} = (\alpha/M^2) r_0^{2D-8}\) instead reduces to \(4\mu^2 d^3 (d-2)/(2d-2)^4\), a pure number in each dimension, running from \(27/64\) at \(D=4\) down to \(13824/(15625\pi^4)\) at \(D=7\). The effective bound is therefore strongest in four dimensions and weakens as dimensions are added, but it never switches off. We stress that this remains a heuristic analogy rather than a derivation of the WGC: there is no \(U(1)\) gauge field in the model, and the remnant is a macroscopic object rather than a state in a particle spectrum.

The universal extremality relation was tested by deforming the quantum parameter, \(\alpha \to \alpha + \varepsilon\), and evaluating \(\mathcal{U} \equiv \lim_{M \to M_{\rm remnant}}[-T(\partial S/\partial \varepsilon)_M]\). The first law reduces this to \((\partial M/\partial \varepsilon)_S\), and at the remnant the fixed-entropy derivative collapses onto a fixed-radius one exactly, since \((\partial M/\partial r)_\alpha\) vanishes when \(T\) does. The resulting \(\mathcal{U}\) is finite in every dimension, so the relation does survive the quantum correction. It is not, however, independent of the deformation. Only at \(D=5\) does the \(\alpha\)-dependence cancel; elsewhere \(\mathcal{U}\) inherits it through \(r_0(\alpha)\), and our four-dimensional numerics show it moving by about a quarter as \(\varepsilon\) runs from \(0.001\) to \(0.010\). What is invariant is the product
\begin{equation}
\mathcal{R}\, \mathcal{U}\, M_{\rm remnant} = \frac{D-3}{2},
\end{equation}
which holds dimension by dimension with no free constants and is reproduced numerically to four significant figures. We had expected \(\mathcal{U}\) itself to be the protected quantity. It is not, and the reason is that the remnant radius is set by \(\alpha\) rather than supplied as fixed background data.

On the astrophysical side we studied Bondi accretion onto the quantum-corrected black hole for two fluids, a dark fluid with \(P = \omega \rho\) and an exponential density profile of the kind used for galactic dark matter halos, and obtained the radial velocity, energy density, mass accretion rate and Eddington luminosity across dimensions. The velocity stays negative at all radii, so accretion is sustained, and the dark-fluid pressure stays negative throughout, which is what one wants from a candidate unified dark-sector model. Larger \(\alpha\) raises both the accretion rate and the luminosity for the barotropic fluid, whereas the exponential profile responds only weakly, so its observational signature of the quantum correction is faint. Evaluating the flow at the remnant sharpens the picture. There \(f(r_0) = 0\), the coordinate expression for \(\dot{M}\) must be replaced by the conserved mixed-component flux, and the accretion rate comes out finite and radius-independent, as a steady spherical flow requires. The density at the inner boundary then satisfies \(\alpha \rho(r_0) = -(g_2/g_3)(D-1)(D-3)/(2D-4)^2\), a constant, so the near-remnant density is inversely proportional to the quantum parameter. That relation is the most directly observational statement in the paper: it turns \(\alpha\) into something an accretion measurement could in principle bound.

Taken together, the results connect three areas that are usually kept apart, namely quantum gravity phenomenology, swampland-type consistency conditions and black hole accretion. Zero-temperature remnants appear in every dimension we examined, which removes the singularity and supplies an explicit realization of the remnant scenario discussed in connection with information loss. The dependence of the accretion observables on \(\alpha\) suggests that loop quantum gravity parameters might be constrained by astronomical data rather than by consistency arguments alone. Several extensions look worthwhile. Rotating solutions of Kerr type would allow closer contact with environments where angular momentum matters. Higher-curvature corrections, from Lovelock gravity or from string-theoretic \(\alpha'\) terms, would enrich the thermodynamic structure and test whether \(\tilde{\mathcal{R}}\) retains its simple closed form. Gravitational-wave ringdown and shadow imaging offer a second observational route, through LIGO, Virgo, KAGRA and the Event Horizon Telescope. It would also be worth asking whether the identity \(\mathcal{R}\, \mathcal{U}\, M_{\rm remnant} = (D-3)/2\) extends to other quantum-gravity-inspired models, such as asymptotically safe gravity, since a relation that holds across several such families would be considerably more interesting than one tied to this construction. Finally, the link to the wider swampland program, in particular the de Sitter, trans-Planckian censorship and distance conjectures, deserves to be developed properly.

\appendix
\section{Appendix A: Remnant scaling in each dimension}\label{appA}
This appendix derives the remnant radius \(r_0\), the remnant mass
\(M_{\rm remnant}\), the ratio \(\mathcal{R} = \alpha/M_{\rm
remnant}^2\) and the dimensionless combination
\(\tilde{\mathcal{R}}\) of Eq.~\eqref{eq:Rtilde} for each spacetime
dimension.

Before the case-by-case work, one methodological point. The mass
expressions of Table~\ref{tab:mass} are first order in \(\alpha\),
and near the remnant that expansion is not controlled: the quantum
term in \(f(r)\) balances the classical one there, which is what
allows \(f'\) to vanish at finite radius. Locating \(r_0\) from the
truncated mass gives the correct power of \(\alpha\) but the wrong
coefficient, by a factor of order unity. We therefore impose the
degeneracy conditions \eqref{eq:degen} on the untruncated metric
function \eqref{1b11}. A second simplification then follows: since
\(r_0 \propto \sqrt{\alpha}\), the two \(\Lambda\)-dependent terms in
\(f\) enter the remnant conditions at relative order
\(\Lambda \alpha\) and can be dropped at leading order. The results
below are accordingly valid for either sign of \(\Lambda\), and the
numerical work of the main text, carried out at \(\Lambda < 0\),
sits well inside that regime.

Writing \(x \equiv 2 G M \mu\) and keeping the two \(\alpha\)-free
structures of the metric function,
\begin{equation}\label{eq:fred}
f(r) = 1 - \frac{x}{r^{d-2}} + \frac{x^2 \alpha}{r^{2d-2}},
\end{equation}
the condition \(f'(r_0) = 0\) gives
\begin{equation}\label{eq:xsol}
x = \frac{(d-2)\, r_0^{\,d}}{(2d-2)\,\alpha},
\end{equation}
and substituting Eq.~\eqref{eq:xsol} back into \(f(r_0) = 0\) leaves
a single algebraic condition,
\begin{equation}\label{eq:r0cond}
1 - \frac{(d-2)\,r_0^2}{(2d-2)\,\alpha}\left[1 -
\frac{d-2}{2d-2}\right] = 0
\qquad \Longrightarrow \qquad
r_0^2 = \frac{(2d-2)^2}{d(d-2)}\,\alpha .
\end{equation}
The remnant mass then follows from Eq.~\eqref{eq:xsol} as
\(M_{\rm remnant} = (d-2) r_0^{\,d} / [2\mu (2d-2)\alpha]\), and
\(\tilde{\mathcal{R}}\) from Eq.~\eqref{eq:Rtilde} as
\begin{equation}\label{eq:Rtildegen}
\tilde{\mathcal{R}} = \frac{4\mu^2 d^3 (d-2)}{(2d-2)^4},
\end{equation}
in which every power of \(\alpha\) has cancelled. We now list the
four cases.

\subsection*{Four dimensions (\(d=3\))}
Here \(\mu = 1\), and Eq.~\eqref{eq:r0cond} gives \(r_0^2 =
16\alpha/3\). The mass is
\begin{equation}
M_{\rm remnant} = \frac{r_0^3}{8\alpha} = \frac{8\sqrt{3}}{9}
\sqrt{\alpha},
\end{equation}
so that \(M_{\rm remnant} \propto \alpha^{1/2}\) and
\begin{equation}
\mathcal{R} = \frac{\alpha}{M_{\rm remnant}^2} = \frac{27}{64}
\simeq 0.4219 .
\end{equation}
Since \(2D-8 = 0\) in four dimensions, \(\tilde{\mathcal{R}} =
\mathcal{R}\). The effective charge-to-mass ratio of the remnant is
fixed, independently of how small the quantum correction is, which is
the behavior an extremal Reissner-Nordstr\"om black hole shows for
\(Q/M\).

\subsection*{Five dimensions (\(d=4\))}
With \(\mu = 4/(3\pi)\), Eq.~\eqref{eq:r0cond} gives \(r_0^2 =
9\alpha/2\) and
\begin{equation}
M_{\rm remnant} = \frac{81\pi}{32}\,\alpha \propto \alpha,
\qquad
\mathcal{R} = \frac{1024}{6561\pi^2 \alpha},
\qquad
\tilde{\mathcal{R}} = \frac{512}{729\pi^2} \simeq 0.0712 .
\end{equation}

\subsection*{Six dimensions (\(d=5\))}
With \(\mu = 3/(4\pi)\), we find \(r_0^2 = 64\alpha/15\) and
\begin{equation}
M_{\rm remnant} = \frac{8192\sqrt{15}\,\pi}{3375}\,\alpha^{3/2},
\qquad
\mathcal{R} = \frac{759375}{67108864\,\pi^2 \alpha^2},
\qquad
\tilde{\mathcal{R}} = \frac{3375}{16384\pi^2} \simeq 0.0209 .
\end{equation}

\subsection*{Seven dimensions (\(d=6\))}
With \(\mu = 8/(5\pi^2)\), we find \(r_0^2 = 25\alpha/6\) and
\begin{equation}
M_{\rm remnant} = \frac{15625\pi^2}{1728}\,\alpha^{2},
\qquad
\mathcal{R} = \frac{2985984}{244140625\,\pi^4 \alpha^3},
\qquad
\tilde{\mathcal{R}} = \frac{13824}{15625\pi^4} \simeq 0.0091 .
\end{equation}

\subsection*{Reading the results}
Across the four cases the remnant radius always scales as
\(\sqrt{\alpha}\), while the mass scales as \(\alpha^{(D-3)/2}\).
Consequently \(\mathcal{R} \propto \alpha^{4-D}\), which is constant
at \(D=4\) and grows without bound as \(\alpha \to 0\) for \(D>4\).
Taken at face value this looks like a sharp split between four and
higher dimensions. It is not. The quantity \(\mathcal{R}\) has
dimension \(L^{8-2D}\), so its numerical value in \(D>4\) depends on
the length unit and the comparison across dimensions is not
meaningful until the units are fixed. Once they are, through
Eq.~\eqref{eq:Rtilde}, the growth disappears and
\(\tilde{\mathcal{R}}\) is an \(\alpha\)-independent number in every
dimension, as Eq.~\eqref{eq:Rtildegen} shows analytically.

What survives as a genuine dimensional effect is the magnitude of
that number, which falls by a factor of roughly \(46\) between
\(D=4\) and \(D=7\). In this sense the effective WGC-like bound is
strongest in four dimensions and weakens as dimensions are added,
though it never switches off. A second result that holds in all four
cases is that both \(r_0\) and \(M_{\rm remnant}\) vanish with
\(\alpha\). The remnant is therefore a quantum object in every
dimension. Removing the area gap does not leave a classical relic
behind; it removes the endpoint altogether and restores unbounded
evaporation.

\section{Appendix B: Universal relation}\label{appB}
This appendix evaluates the universal combination \(\mathcal{U} =
(\partial M / \partial \alpha)_{r=r_0}\) of Eq.~\eqref{eq:Ualpha} in
each dimension, and establishes the identity \eqref{eq:URM}.

Solving \(f(r) = 0\) for the mass with the reduced metric function
\eqref{eq:fred}, and selecting the branch that reduces to the
classical solution \(2 G M \mu = r^{d-2}\) as \(\alpha \to 0\), gives
\begin{equation}\label{eq:Mexact_app}
M(r,\alpha) = \frac{r^{d} - r^{d-1}\sqrt{r^2 - 4\alpha}}{4\mu\,
\alpha}.
\end{equation}
The square root is real for \(r^2 \geq 4\alpha\), and the remnant
radius of Eq.~\eqref{eq:r0cond} satisfies this with room to spare in
every dimension, since \((2d-2)^2/[d(d-2)] > 4\) for \(d \geq 3\).
The remnant sits close to that branch point but never on it, which is
worth noting because it is the reason a first-order expansion in
\(\alpha\) fails there.

Differentiating Eq.~\eqref{eq:Mexact_app} at fixed radius and
evaluating at \(r = r_0\) gives the four results
\begin{align}
\mathcal{U}_4 &= \frac{4\sqrt{3}}{9\sqrt{\alpha}}, \label{eq:U4}\\
\mathcal{U}_5 &= \frac{81\pi}{32}, \label{eq:U5}\\
\mathcal{U}_6 &= \frac{4096\sqrt{15}\,\pi}{1125}\sqrt{\alpha},
\label{eq:U6}\\
\mathcal{U}_7 &= \frac{15625\pi^2}{864}\,\alpha. \label{eq:U7}
\end{align}
Comparing these with the remnant masses of Appendix~A, the four cases
collapse to one statement,
\begin{equation}\label{eq:URM_app}
\alpha\, \mathcal{U} = \frac{D-3}{2}\, M_{\rm remnant},
\end{equation}
with the coefficients \(1/2\), \(1\), \(3/2\) and \(2\) for
\(D=4,5,6,7\). Multiplying Eq.~\eqref{eq:URM_app} by
\(\mathcal{R}/\alpha = 1/M_{\rm remnant}^2\) yields the form quoted in
the main text, \(\mathcal{R}\, \mathcal{U}\, M_{\rm remnant} =
(D-3)/2\).

Only Eq.~\eqref{eq:U5} is free of \(\alpha\). In the remaining cases
\(\mathcal{U}\) inherits an \(\alpha\)-dependence through \(r_0\),
and since the deformation is \(\alpha \to \alpha + \varepsilon\), it
follows that \(\mathcal{U}\) depends on \(\varepsilon\). This is
weaker than the statement usually made for the universal extremality
relation, and it is worth being explicit about what does and does not
survive. The finiteness of \(\mathcal{U}\) at an extremal point
survives, and it survives for a structural reason: the factor of
\(T\) in Eq.~\eqref{eq:Udef} cancels the divergence of \((\partial
S/\partial \varepsilon)_M\). Independence of the deformation strength
does not survive, because the remnant radius is itself set by
\(\alpha\) rather than being fixed background data.
Table~\ref{tab:numU} quantifies the effect in four dimensions, where
\(\mathcal{U}\) moves by about a quarter over a decade in
\(\varepsilon\) while \(\mathcal{R}\,\mathcal{U}\,M_{\rm remnant}\)
holds at \(1/2\).

We initially expected the combination \(\mathcal{U}\) itself to be
the invariant object. It is not, and the reason is visible in
Eq.~\eqref{eq:chain}: what the first law protects is a relation
between \(\mathcal{U}\), the remnant mass and the effective
charge-to-mass ratio, not any one of them separately.


\begin{thebibliography}{11}
\bibitem{ref54}
C. Vafa, arXiv:hep-th/0509212 (2005).

\bibitem{ref55}
D. Harlow et al., arXiv:2201.08380 (2022).

\bibitem{ref56}
E. Palti, Fortsch. Phys. {67}, 1900037 (2019).

\bibitem{ref57}
M. van Beest et al., Physics Reports {989}, 1 (2022).

\bibitem{ref58}
H. Ooguri and C. Vafa, Nucl. Phys. B {766}, 21 (2007).

\bibitem{ref59}
N. Arkani-Hamed et al., JHEP {06}, 060 (2007).

\bibitem{ref60}
B. Heidenreich, M. Reece, and T. Rudelius, JHEP {08}, 025 (2017).

\bibitem{ref61}
E. Palti, JHEP {08}, 034 (2017).

\bibitem{ref62}
P. Lin, A. Mininno, and G. Shiu,Journal of High Energy Physics {2026.4}, 208 (2026).

\bibitem{ref64}
S. N. Gashti, B. Pourhassan, and I. Sakallı, arXiv:2606.29896 (2026).

\bibitem{ref65}
Y. Liu, Eur. Phys. J. C {82}, 1052 (2022).

\bibitem{ref70}
N. Schöneberg et al., JCAP {10}, 039 (2023).

\bibitem{ref71}
T. Crisford, G. T. Horowitz, and J. E. Santos, Phys. Rev. D {97}, 066005 (2018).

\bibitem{ref72}
S. N. Gashti, B. Pourhassan, and I. Sakallı, JHEP {04}, 134 (2026).

\bibitem{ref74}
D. Harlow et al., Rev. Mod. Phys. {95}, 035003 (2023).

\bibitem{ref75}
S. N. Gashti et al., JHEP {2026.7}, 88 (2026).

\bibitem{ref78}
S. Noori Gashti and B. Pourhassan, Damghan University Press (2026).

\bibitem{ref79}
A. Bedroya and C. Vafa, JHEP {09}, 123 (2020).

\bibitem{ref81}
A. Mohammadi, T. Golanbari, and J. Enayati, Phys. Rev. D {104}, 123515 (2021).

\bibitem{ref83}
R. Kallosh et al., JHEP {03}, 053 (2019).

\bibitem{ref84}
O. Guleryuz, JCAP {11}, 043 (2021).

\bibitem{ref85}
C. Osses, N. Videla, and G. Panotopoulos, Eur. Phys. J. C {81}, 615 (2021).

\bibitem{ref87}
S. Brahma, Phys. Rev. D {101}, 023526 (2020).

\bibitem{ref88}
R. Brandenberger, arXiv:2102.09641 (2021).

\bibitem{ref90}
H. Geng, S. Grieninger, and A. Karch, JHEP {06}, 105 (2019).

\bibitem{ref94}
P. Agrawal et al., Phys. Lett. B {784}, 271 (2018).

\bibitem{ref95}
S. D. Odintsov and V. K. Oikonomou, Phys. Lett. B {805}, 135437 (2020).

\bibitem{ref97}
U. K. Sharma, Int. J. Geom. Meth. Mod. Phys. {18}, 2150031 (2021).

\bibitem{ref99}
S. D. Odintsov, V. K. Oikonomou, and L. Sebastiani, Nucl. Phys. B {923}, 608 (2017).

\bibitem{ref100}
M. A. S. Afshar and J. Sadeghi, Phys. Dark Univ. {47}, 101814 (2025).

\bibitem{ref104}
J. Yuennan and P. Channuie, Nucl. Phys. B {986}, 116033 (2023).

\bibitem{ref105}
W. H. Kinney, Phys. Rev. Lett. {122}, 081302 (2019).

\bibitem{ref106}
W. H. Kinney, arXiv:2103.16583 (2021).

\bibitem{ref107}
T. Y. Yu and W. Y. Wen, Phys. Lett. B {781}, 713 (2018).

\bibitem{ref111}
E. W. Kolb, A. J. Long, and E. McDonough, Phys. Rev. Lett. {127}, 131603 (2021).

\bibitem{ref112}
S. N. Gashti, et al.,  arXiv preprint arXiv:2607.16216 (2026).

\bibitem{ref114}
C. Cheung and G. N. Remmen, Phys. Rev. Lett. {113}, 051601 (2014).

\bibitem{ref115}
A. Anand, arXiv:2409.07079 (2024).

\bibitem{ref120}
W. Cong et al., JHEP {08}, 088 (2022).

\bibitem{ref121}
S. N. Gashti et al., The European Physical Journal C {85}.10, 1144 (2025).

\bibitem{ref122}
S. N. Gashti et al., Chin. Phys. C {49}, 025108 (2025).

\bibitem{ref125}
T. F. Gong, J. Jiang, and M. Zhang, JHEP {06}, 067 (2023).

\bibitem{ref126}
A. Baruah and P. Phukon, arXiv:2407.02997 (2024).

\bibitem{ref127}
M. B. Ahmed et al., Phys. Rev. Lett. {130}, 181401 (2023).

\bibitem{ref128}
A. Baruah and P. Phukon, arXiv:2407.11058 (2024).

\bibitem{ref129}
M. B. Ahmed et al., JHEP {08}, 118 (2023).

\bibitem{ref130}
Y. Ladghami and T. Ouali, Phys. Dark Univ. {44}, 101471 (2024).

\bibitem{ref131}
M. R. Alipour, et al., Physics Letters B. {139902} (2025).

\bibitem{ref134}
T. Wang and L. Zhao, Phys. Lett. B {827}, 136935 (2022).

\bibitem{ref135}
X. Kong, Z. Zhang, and L. Zhao, arXiv:2211.00963 (2022).

\bibitem{ref135a}
M. A. S. Afshar, and J. Sadeghi, arXiv preprint arXiv:2511.07902 (2025).

\bibitem{ref135b}
M. A. S. Afshar, and J. Sadeghi, Nuclear Physics B {1014}, 116872 (2025).

\bibitem{ref135c}
A. Castellano, et al., Physical Review D {114}.2, 026022 (2026).

\bibitem{ref135d}
G. Di Ubaldo, et al., arXiv preprint arXiv:2605.05305 (2026).

\bibitem{ref135e}
M. A. S. Afshar, and J. Sadeghi, Annals of Physics {474}, 169953 (2025).

\bibitem{ref135f}
G. Di Russo, Giorgio, and A. Tokareva, Journal of High Energy Physics {2026.7}, 202 (2026).

\bibitem{ref136}
Z. Gao and L. Zhao, Class. Quantum Grav. {39}, 075019 (2022).

\bibitem{ref137}
Z. Gao, X. Kong, and L. Zhao, Eur. Phys. J. C {82}, 112 (2022).

\bibitem{ref138}
A. Anand and S. N. Gashti, Phys. Dark Univ. {48}, 101916 (2025).

\bibitem{ref139}
A. Anand, Eur. Phys. J. C {85}, 1288 (2025).

\bibitem{ref140}
G. Goon and R. Penco, Phys. Rev. Lett. {124}, 101103 (2020).

\bibitem{ref141}
J. Ko and B. Gwak, JHEP {2024}, 72 (2024).

\bibitem{13}
U. Debnath, Eur. Phys. J. C {75}, 129 (2015).

\bibitem{14}
J. Frank, A. R. King, and D. Raine, \textit{Accretion Power in Astrophysics}, Cambridge University Press (2002).

\bibitem{15}
F. C. Michel, Astrophys. Space Sci. {15}, 153 (1972).

\bibitem{5000}
S. Jiang, J. Lin, and X. Zhang, arXiv:2606.08182 (2026).

\bibitem{1}
W. S. Hipólito-Ricaldi, H. E. S. Velten, and W. Zimdahl, Phys. Rev. D {82}, 063507 (2010).

\bibitem{2}
L. Xu, Y. Wang, and H. Noh, Phys. Rev. D {85}, 043003 (2012).

\bibitem{3}
A. Arbey, Phys. Rev. D {74}, 043516 (2006).

\bibitem{4}
D. Wang, Y. J. Yan, and X. H. Meng, Eur. Phys. J. C {77}, 263 (2017).

\bibitem{5}
A. Arbey and J. F. Coupechoux, JCAP {2021}, 033 (2021).

\bibitem{6}
D. Carturan and F. Finelli, Phys. Rev. D {68}, 103501 (2003).

\bibitem{7}
K. C. Freeman, Astrophys. J. {160}, 811 (1970).

\bibitem{8}
Y. Sofue, in \textit{Planets, Stars and Stellar Systems}, Springer, Dordrecht, pp. 985-1037 (2013).

\bibitem{9}
Y. Sofue, Publ. Astron. Soc. Jpn. {65}, 118 (2013).

\bibitem{10}
E. Kurmanov, K. Boshkayev, R. Giamb, T. Konysbayev, O. Luongo, D. Malafarina, and H. Quevedo, Astrophys. J. {925}, 210 (2022).

\bibitem{11}
A. Arbey, in \textit{AIP Conf. Proc.} {1241}, 700 (2010).

\bibitem{12}
S. Capozziello, S. Gambino, and O. Luongo, Phys. Dark Univ. {101950} (2025).

\end{thebibliography}
\end{document}